\documentclass[superscriptaddress,nofootinbib,preprintnumbers,amsmath,amssymb, preprint]{revtex4-1}

\usepackage[T1]{fontenc}

\usepackage[titles]{tocloft}
\cftsetindents{section}{0em}{2.5em}
\cftsetindents{subsection}{2.5em}{2.5em}

\usepackage{multirow}
\usepackage{tabularx}
\usepackage{amsfonts}
\usepackage{mathrsfs}
\usepackage{leftidx}
\usepackage{amsmath}
\usepackage{amssymb}
\usepackage{placeins}
\usepackage{relsize}
\usepackage{slashed}

\usepackage[dvipsnames]{xcolor}
\definecolor{red}{rgb}{0.9, 0,0}
\definecolor{cerulean}{rgb}{0., 0.42,0.9}
\definecolor{navy}{rgb}{0.05, 0.05,0.8}

\usepackage[colorlinks]{hyperref}
\hypersetup{
     colorlinks   = true,
     citecolor    = red,
	linkcolor = navy
}

\usepackage{soul}
\usepackage{epsfig}
\usepackage{graphicx}               
\usepackage{url}
\usepackage{float}
\usepackage{soul}

\newcommand{\be}{\begin{equation}}
\newcommand{\ee}{\end{equation}}
\newcommand{\bea}{\begin{eqnarray}}
\newcommand{\eea}{\end{eqnarray}}
\newcommand{\beq}{\begin{eqnarray}}
\newcommand{\eeq}{\end{eqnarray}}
\def\bit{\begin{itemize}}
\def\eit{\end{itemize}}
\def\ben{\begin{enumerate}}
\def\een{\end{enumerate}}

\newcommand{\Eqs}[2]{Eqs.~(\ref{#1}) and (\ref{#2})}

\def\lag{\mathcal{L}}

\newcommand{\overbar}[1]{\mkern 2.5mu\overline{\mkern-2.5mu#1\mkern-2.5mu}\mkern 2.5mu}

\def\lag{\mathcal{L}}

\def\mxbar{{\overbar{m}_\chi}}

\newcommand{\OO}{\mathcal{O}}

\newcommand{\bfq}{{\bf q}}

\newcommand{\bfv}{{\bf v}}
\newcommand{\bfr}{{\bf r}}

\newcommand{\bra}{\langle}
\newcommand{\ket}{\rangle}

\newcommand\DN[1][\relax]{%
\ifx\relax#1\relax\else{}^{#1}\fi \!X}

\newcommand{\GeV}{{\,\rm GeV}}

\newcommand{\keV}{{\,\rm keV}}
\newcommand{\eV}{{\,\rm eV}}

\newcommand{\MeV}{{\,\rm MeV}}

\DeclareMathAlphabet\mathbfcal{OMS}{cmsy}{b}{n} 

\graphicspath{{./figures/}}

\begin{document}


\title{Direct Detection of Bound States of Asymmetric Dark Matter}

\author{Ahmet Coskuner}
\affiliation{Theoretical Physics Group, Lawrence Berkeley National Laboratory, Berkeley, CA 94720}
\affiliation{Berkeley Center for Theoretical Physics, University of California, Berkeley, CA 94720}
\author{Dorota M Grabowska}
\affiliation{Berkeley Center for Theoretical Physics, University of California, Berkeley, CA 94720}
\affiliation{Theoretical Physics Group, Lawrence Berkeley National Laboratory, Berkeley, CA 94720}
\author{Simon Knapen}
\affiliation{School of Natural Sciences, Institute for Advanced Study, Princeton, NJ 08540, USA}
\affiliation{Theoretical Physics Group, Lawrence Berkeley National Laboratory, Berkeley, CA 94720}
\affiliation{Berkeley Center for Theoretical Physics, University of California, Berkeley, CA 94720}
\author{Kathryn M. Zurek}
\affiliation{Theoretical Physics Group, Lawrence Berkeley National Laboratory, Berkeley, CA 94720}
\affiliation{Berkeley Center for Theoretical Physics, University of California, Berkeley, CA 94720}
\affiliation{Theory Department, CERN, Geneva, Switzerland}

\begin{abstract} 
We study the reach of direct detection experiments for large bound states (containing $10^4$ or more dark nucleons) of Asymmetric Dark Matter.  We consider ordinary nuclear recoils, excitation of collective modes (phonons), and electronic excitations, paying careful attention to the impact of the energy threshold of the experiment.  Large exposure experiments with keV energy thresholds provide the best (future) limits when the Dark Matter is small enough to be treated as a point particle, but rapidly lose sensitivity for more extended dark bound states, or when the mediator is light. In those cases, low threshold, low exposure experiments (such as with a superfluid helium, polar material or superconducting target) are often more sensitive due to coherent enhancement over the dark nucleons. We also discuss indirect constraints on composite Asymmetric Dark Matter arising from self-interaction, formation history and the properties of the composite states themselves. 
\end{abstract}

\maketitle

\tableofcontents
\clearpage


\section{Introduction}

The last decade has seen a dramatic broadening in the types of Dark Matter (DM) theories that are being proposed and searched for using various experiments \cite{Alexander:2016aln,Battaglieri:2017aum}.  Previously the Weakly Interacting Massive Particle (WIMP) and the axion were the focus of both theoretical and experimental attention, and for good reason: in addition to solving the DM mystery, they resolve another theoretical puzzle, the hierarchy problem and the strong CP problem, respectively.  As the WIMP has not surfaced, the urgency to look elsewhere has increased.  At the same time, a qualitative expansion of the number of DM candidates has occurred.  Most of the new ideas have centered around {\em hidden sector} DM, where the dynamics of the non-Standard Model sector allow for a wider range of dark matter candidates and signatures.  


Most of the work in hidden sector dark matter has focused on candidates of a low mass \cite{ Boehm:2003hm,  Strassler:2006im, Pospelov:2008jd, Feng:2008ya, Zurek:2013wia, Hochberg:2014dra}, as envisioned and proposed in the hidden valley model.  This focus is partly phenomenologically motivated, but as thermal candidates with mass above 10 TeV cannot obtain the correct relic abundance through a standard thermal freeze-out scenario, there is also theoretical motivation.  A natural question is what models of DM exist above this ``unitarity bound'' and how one searches for such candidates.

Here we consider the detection of very heavy composite bound states of Asymmetric Dark Matter (ADM).  These states satisfy the unitarity bound because they are synthesized relatively late in the Universe from light constituents in a hidden sector.  The symmetric component of the DM freezes out through annihilation to light force mediators in the hidden sector, as proposed in the original models \cite{Kaplan:2009ag,Lin:2011gj}.  If the forces in the hidden sector are sufficiently strongly attractive and long range, the DM states will bind and grow into large states, as shown in \cite{Wise:2014ola,Wise:2014jva, Krnjaic:2014xza, Hardy:2014mqa,Detmold:2014qqa}.  When the states grow to a size such that the Fermi degeneracy pressure dominates the dynamics of the bound state, they ``saturate'' to a constant density \cite{Gresham:2017zqi,Gresham:2017cvl}; we call these states ADM nuggets. If the confining force in the dark sector undergoes a first order phase transition, dark quark nuggets can form\cite{Witten:1984rs,Bai:2018dxf}, though in this paper we focus on nuggets formed via fusion. Because the properties of ADM nuggets depend on very few parameters, such as the force range and strength and the constituent masses, and combined with the requirement that the states can form in the first place, the theory space is fairly predictive.  It was found in \cite{Gresham:2018anj} that the natural size of ADM nuggets, formed via early universe fusion, is $M_X \lesssim 10^{20} \mbox{ GeV}$, and that they give distinctive signatures in the structure of galaxies and can produce signatures in indirect detection \cite{Detmold:2014qqa,Gresham:2018anj}.

Due to their large size, the scattering of ADM nuggets in direct detection experiments can benefit from a  significant $N_X^2$ coherent enhancement, with $N_X$ the number of constituents in the nugget.  The corresponding phenomenology was studied in~\cite{Hardy:2015boa,Butcher:2016hic}, where it was shown how the recoil spectrum of an ADM nugget can be distinguished from that of normal WIMP DM. The focus of this previous work was primarily on noble liquid and semi-conductor experiments, and in this paper we aim to extend it by including more recent, lower threshold ideas for both electron and nucleon couplings. We furthermore consider both massive and light mediators, as well as a light, kinetically mixed dark photon mediator. Our motivation is that the $N_X^2$ enhancement only holds \emph{as long  the momentum transfer is low compared to the inverse radius of the nugget}, which correlates strongly with the experimental threshold. This effect was recently exploited for the largest composite objects, $M_X \gtrsim 10^{20} \mbox{ GeV}$, whose interaction cross-section is dominated by the object's geometric size, in the context in particular of gravitational wave detectors and spin precession experiments \cite{Grabowska:2018lnd}. Here we focus on the regime most motivated by synthesis considerations (in the absence of a bottleneck, see \cite{Gresham:2018anj}), $10^3 \mbox{ GeV}\lesssim M_X \lesssim 10^{20} \mbox{ GeV}$, and show how the recent program towards ultra low threshold detectors can impact the nugget DM parameter space.

 In particular, a number of experiments have been recently proposed that are sensitive to very low momentum transfers, both for DM coupling to nucleons (see e.g.~\cite{Essig:2016crl,Kadribasic:2017obi,Budnik:2017sbu,Schutz:2016tid,Knapen:2016cue,Knapen:2017ekk,Griffin:2018bjn}) as well as electrons (see e.g.~\cite{Essig:2011nj,Essig:2015cda,Hochberg:2016ntt,Derenzo:2016fse,Hochberg:2015pha,Hochberg:2017wce}); some of these experiments are also sensitive to interactions via dark photons.   Due to the increased sensitivity via the coherent enhancement, these experiments are natural places to look when searching for ADM nuggets.  However, these novel experiments are very small in volume compared to the large noble liquid detectors used in classic WIMP searches.  Our goal is to quantify the relative reach of low threshold experiments compared to the classic WIMP-like searches.  We will show that the relative sensitivity at a given nugget mass depends strongly on the constituent mass, which (largely) fixes the nugget radius.  For lower constituent masses, where the nugget saturation densities are lower and their size is larger, the low threshold experiments dominate; for larger constituent masses, the opposite is true and the noble-liquid experiments provide better sensitivity. This result highlights the necessity of a multi-prong experimental approach to place constraints on ADM nuggets.

The outline of this paper is as follows.  In the Sec.~\ref{sec:CompositeDM} we review the model of composite dark matter we employ, and lay out our conventions and the main formulas for the direct detection of nuggets. We also review the relevant constraints from probes other than direct detection.  In Secs.~\ref{sec:nucleoncoupling} and \ref{sec:electroncoupling} we consider the direct detection prospects of ADM nuggets with couplings to nucleons and electrons respectively. We conclude in Sec.~\ref{sec:conclusions}.

\section{Composite Asymmetric Dark Matter}
\label{sec:CompositeDM}
The goal of this section is to give a brief review of ADM nugget properties and their formation history. This allows us to select a natural model space for direct detection, as well as define our conventions for the direct detection parameter space. We also discuss the relevant constraints on the particle mediating the scattering with the Standard Model (SM), from probes other than direct detection.

\subsection{Model and formation history}

In the absence of a bottleneck, such as provided by electromagnetism in the SM, fermionic DM with a sufficiently attractive and long-range force can be synthesized with as many as $\sim 10^{20}$ constituents.  We refer the reader to \cite{Gresham:2018anj} for a review of nugget synthesis and properties.   A generic Lagrangian to describe the properties of bound states of the fermions is given by
\begin{align}
  \mathcal{L} = \bar \chi(i\slashed\partial - m_\chi)\chi - \frac{1}{2} m_\varphi^2\varphi^2
 + \frac{1}{2} m_V^2V_\mu^2
- \bar \chi \left[ g_\varphi \varphi + g_V \slashed V \right]\chi
- V(\varphi, a, V, A) \, ,
\label{eq:lagrangian}
\end{align}
where $\chi$ are the constituent fermions of the bound state and  for compactness we have omitted the kinetic terms of the scalar, $\varphi$, and vector, $V$, mediators.  Such a weakly coupled model (with $\varphi$ relabeled as the $\sigma$ and $V$ relabeled as the $\omega$) is employed in nuclear physics to describe the gross features of the underlying strong dynamics.

To determine the reach of direct detection experiments, we need to parametrize the properties of the nuggets. The properties of nuggets with only a few constituents are quite model-dependent. However, once the nugget radius exceeds the de Broglie wavelength of the force mediator, the nugget enters a constant density regime called saturation.  Here, it can be shown that all the nugget properties can be parameterized in terms of just two quantities \cite{Gresham:2017cvl,Gresham:2018anj} -- the constituent mass $m_\chi$ and the reduced constituent mass $\mxbar$, with the latter taking into account the in-medium effects of the nugget on the constituent mass.  For example, the radius of a bound state of $N_X$ constituents,
\beq
R_X = \left(\frac{N_X}{4/3 \pi n_{\rm sat}}\right)^{1/3},
\label{eq:R}
\eeq
is determined by the saturation density $n_{\rm sat}$.  This quantity is in turn is set by $\mxbar$:   
\begin{align}
  \frac{n_{\rm sat}}{\bar{m}_\chi^3} =
                \begin{cases}
                  \frac{1}{3\pi^2} & \frac{C_V^4}{C_{\varphi}^{2}} \le \frac{1}{8} \\
                  \frac{1}{3\pi^2}\left[\frac{1}{2}+ \left(\frac{C_V^4}{C_{\varphi}^2}\right)^{\frac{1}{3}}\right]^{-3} & \frac{C_V^4}{C_{\varphi}^{2}} > \frac{1}{8}
                \end{cases}
\qquad \quad
\frac{\mxbar}{m_\chi}=
                    \begin{cases}
                      \left(\frac{2}{C_{\varphi}^2} \right)^{\frac{1}{4}}
                      & \frac{C_V^4}{C_{\varphi}^{2}} \le \frac{1}{8} \\
                       \frac{1}{(C_\varphi C_V)^{\frac{1}{3}}}\left[\frac{1}{2}+ \left(\frac{C_V^4}{C_{\varphi}^2}\right)^{\frac{1}{3}}\right] & \frac{C_V^4}{C_{\varphi}^{2}} > \frac{1}{8}                 
                    \end{cases} \label{eq: nugget params}
\end{align}
where
\begin{align}
C_V^2 \equiv {g_V^2 \over 3 \pi^2}{m_\chi^2 \over m_V^2} \qquad \text{and} \qquad C_{\varphi}^{2} \equiv  {g_\varphi^2 \over 3 \pi^2 }\frac{m_\chi^2}{m_\varphi^2}\left[ 1 + \frac{2 g_\varphi^2 V(m_\chi/g_\varphi)}{m_\varphi^2 m_\chi^2} \right]^{-1}\, . \label{eq: CV Cphi definitions}
\end{align}
In order for the bound states to reach saturation, neither $m_\varphi$ or $m_V$ can be too small, i.e, it must be that both $m_\phi$ and $m_V$ must be larger than $R_X^{-1}$.  Thus we see that $m_\chi$ and $\mxbar$ tend not to be widely separated.

The total nugget mass is
\beq
M_X = N_X \mxbar + \epsilon_{\rm surf} N_X^{2/3},
\eeq 
where $\epsilon_{\rm surf}$ is the surface energy density.  In a general model with both attractive and repulsive forces, $\epsilon_{\rm surf}$ is typically within an ${\cal O}(1)$ number of $\mxbar$ and hence is a negligible contribution to the total mass for large nuggets (see Refs.~\cite{Gresham:2017cvl,Gresham:2018anj} for a discussion and calculations).

For our purposes, we choose $M_X = N_X \mxbar$ and $n_{\rm sat} = \bar{m}_\chi^3/(3 \pi^2)$ to illustrate the reach of direct detection experiments.  All our results will be parametrized in terms of $M_X$ and $\mxbar$ with the radius of the nugget in Eq.~\eqref{eq:R} written in terms of these variables,
\beq
R_X = \ \left(\frac{9\pi}{4}\frac{M_X}{\bar{m}_\chi^4}\right)^{1/3}.
\label{eq:radiusformula}
\eeq
Note that throughout this paper we use a convention where the index $\chi$ is used as a constituent label, while $X$ refers to the bound state of $\chi$'s.

A natural question to ask is the typical size of nuggets that are synthesized via early universe fusion.  While the size of synthesized states is in general model-dependent, large nuggets that constitute a significant fraction of the DM energy density are most naturally synthesized in the absence of a bottleneck to formation.  In this case, the dynamics of synthesis are quite model-independent, determined only by the Hubble parameter and the geometric cross section of the nugget.  One finds the synthesized size to be \cite{Gresham:2018anj,Gresham:2018anj}
\beq
N_X \simeq 10^{12} \left(\frac{g_*(T_{\rm syn})}{10}\right)^{3/5} \left(\frac{1\mbox{ GeV}}{\mxbar}\right)^{12/5}\left(\frac{\bar{m}_\chi^3}{n_{\rm sat}}\right)^{4/5}\left(\frac{T_{\rm syn}}{\mxbar}\right)^{9/5}.
\label{eq:N}
\eeq
Thus we see that lower constituent masses tend to give rise to larger nuggets.  The reason for this can be seen in Eq.~\eqref{eq: nugget params}:  nuggets with low constituent masses tend to have lower saturation densities, and hence larger sizes, implying that freeze-out of the synthesis process happens later.  As we will see in the next section, however, self-interaction constraints tend to favor smaller interaction cross sections, implying that nuggets typically are not too large if they compose all the dark matter. Furthermore, if there is an additional mediator, either scalar or vector, whose de Broglie wavelength exceeds the radius of the nugget, its couplings must be extremely small in order to not affect the formation history and structure of ADM nuggets. In this work we will however take the nugget mass as a free parameter, with all nuggets having the same mass for simplicity. We leave the interplay between the nugget mass distribution, as predicted from the formation history, and the direct detection bounds for future studies.

\subsection{Constraints on the mediator}
\label{sec:otherconstraints}

We assume that the Dark Sector interacts with the Standard Model via a mediator\footnote{The mediator $\phi$, responsible for the interactions with the SM, is in general not the same field as the mediator which binds the ADM nugget in Eq.~\eqref{eq:lagrangian}, denoted by $\varphi$.}, $\phi$, which couples either to nucleons or electrons,
\beq
\lag \supset \tfrac{1}{2}m_\phi^2 \phi^2 + g_\chi \phi \, \bar \chi \chi + g_n \phi\, \bar n n + g_e \phi \, \bar e e\,\,\,.
\eeq
The mass of the mediator, $m_\phi$, can either be large or small, compared to the typical momentum transfer in direct detection experiments. In our analysis, we will always consider $g_n$ or $g_e$ separately, in Secs.~\ref{sec:nucleoncoupling} and \ref{sec:electroncoupling} respectively. We will discuss these types of constraints in the rest of this subsection.

\subsubsection{Mediator Interactions with Standard Model}
\label{sec:MedInteractionSM}
The constraints on the coupling of the mediator to the Standard Model are highly dependent on the mediator mass, and we only consider a few benchmark scenarios, corresponding to the heavy and light mediator scenarios. 

For a heavy mediator with a coupling to nucleons, the collider constraints depend on the UV completion of the $g_n \phi\, \bar n n $ coupling. We consider the 
\begin{equation}\label{eq:uvcomp}
\mathcal{L} \supset \frac{\alpha_s}{4\Lambda} \phi G^a G^a 
\end{equation}
operator, where we fix $\Lambda=2$ TeV. $\Lambda$ is approximately the scale where Eq.~\eqref{eq:uvcomp} in turn must be UV completed, and its value is taken such that the UV completion can plausibly be outside the reach of the LHC, depending on the details of the model.  Following \cite{Gunion:1989we}, we then map this coupling to the effective nucleon coupling $g_n= \frac{2\pi}{9}\frac{m_n}{\Lambda}\approx 3.4\times 10^{-4}$ and choose $m_\phi=10$ GeV and $g_\chi=1$. The mediator $\phi$ will then decay invisibly if $2m_\chi< m_\phi$, and to a pair of soft jets otherwise. The collider limits in these cases are extremely weak or non-existing respectively (see e.g.~\cite{Aaboud:2017phn,Sirunyan:2017dnz}), and do not constrain the model for our choice of $\Lambda$. For the above parameter choices we can then estimate an upper bound on the nugget direct detection cross section of
\begin{equation}\label{eq:colliderbound}
\sigma_{Xn} \approx N_X^2\frac{ g_\chi^2g_n^2}{4\pi }\frac{ \mu_{{Xn}}^2}{ m_\phi^4}  \lesssim 3 \times 10^{-34}\times \left(\frac{N_X}{10^3}\right)^2\times \text{cm}^2 
\end{equation}
with $\mu_{{nX}}\approx m_n$ the nugget-nucleon reduced mass. It is possible to further relax this constraint by considering a scenario where $\phi$ has $\mathcal{O}(10^{-1})$ couplings to the up and/or down quark, though this requires a relatively complex flavor model.

For a heavy mediator coupling to electrons, we take $m_\phi=25$ MeV and $g_e=10^{-4}$. This benchmark point is allowed by the current the $(g-2)_e$ and SLAC beam dump bounds \cite{Liu:2016qwd}, as well as the Babar bound \cite{Lees:2014xha,Essig:2013vha} and the projections for Belle II \cite{Kou:2018nap}, both if $\phi$ decays visibly and invisibly. For $2m_\chi<m_\phi$ the mediator can decay to the dark matter, in which case the LSND bound considered in \cite{deNiverville:2011it} applies. With the parameter choices above, it suffices to take $g_\chi=10^{-1}$ to evade this constraint. For these choices, the direct detection cross section is roughly
\begin{equation}\label{eq:lepbound}
\sigma_{Xe} \approx N_X^2\frac{ g_\chi^2g_e^2}{4\pi }\frac{ \mu_{{Xe}}^2}{ m_\phi^4}  \lesssim 2 \times 10^{-33}\times \left(\frac{N_X}{10^3}\right)^2\times \text{cm}^2 .
\end{equation}
We will see in Secs.~\ref{sec:nucleoncoupling} and \ref{sec:electroncoupling} that this bound is largely irrelevant compared to the direct detection constraints, especially for $N_X\gg1$. 

For mediators with eV-scale masses, the most stringent constraints arise from astrophysical considerations, such as stellar cooling constraints. For even lighter mediators, constraints from fifth force and weak equivalence principle violation tests become more important. The constraints on light mediators in the mass range of an eV to a GeV are summarized in e.g.~\cite{Knapen:2017xzo}. In this work, we do not scan over mediator masses, but instead choose a benchmark of $m_\phi=1$ eV to illustrate the feasibility of detecting ADM nuggets via this portal. Stellar cooling constrains $g_n \lesssim 10^{-12}$ and $g_e \lesssim 10^{-15}$ for this mass \cite{Hardy:2016kme}.  Sec.~\ref{sec:electroncoupling} we will also briefly consider a dark photon mediator. The stellar bounds on the mixing parameter $\kappa$ with the SM photon is roughly $\kappa\lesssim10^{-12}\times \left(\frac{m_{A'}}{\mathrm{eV}}\right)$, with $m_{A'}$ the dark photon mass \cite{An:2013yfc}.

\subsubsection{Mediator Interactions with Dark Sector}
\label{sec:MedInteractionDarkSector}
The coupling of the mediator to the dark sector is constrained by DM self-interactions, both in the early universe and in the collisions of galaxy clusters, and by the properties of the bound states in the presence of an additional long-range mediator. 

Self-interaction constraints tend to be quite stringent, as the DM-DM cross section is enhanced by the $N_X^2$ coherence factor. For a heavy (short range) mediator, we can assume that the DM-DM cross section is saturated to the geometric cross section. Since the short range force which binds the ADM nugget also contributes to DM-DM scattering, this remains true regardless of the strength of the DM-SM mediator. The bound is then
\begin{equation}
\sigma^\text{geo}_{XX}=4\pi R_X^2 \lesssim 1\, \text{cm}^2 \times \frac{M_X}{\text{g}},
\end{equation}
where the cross-section is chosen to be consistent with DM dynamics in the Bullet Cluster~\cite{Spergel:1999mh, Kahlhoefer:2013dca}. This can be written as a lower bound on $M_X$ using the relation in Eq.~\eqref{eq:radiusformula}
\begin{equation}
M_X \gtrsim 4 \times 10^{14}\GeV \, \times \,\left(\frac{ \MeV }{\mxbar}\right)^8\, .
\label{eq:GeoSIDM}
\end{equation}
This extreme sensitivity to $\mxbar$ implies that no self-interaction bound exists for $\mxbar\gtrsim 100$ MeV. 

 In the case of a light mediator, the cross section receives a contribution from the corresponding long range interaction, in addition to the purely geometric contribution that arises from the short range force that binds the ADM nugget.  The range of the interaction is therefore larger than the radius of the ADM nugget, but the scattering is biased towards small angles. Therefore, the cross section of interest is the momentum transfer cross section, $\sigma_T$ \cite{Feng:2009mn,McDermott:2010pa}, which can be approximated in terms of the classical "point of closest approach" $d_C$, 
 \beq
\sigma^\text{light}_T \sim 4 \pi \,d^2_C \sim \frac{4 \pi}{m_\phi^2}  \text{W}\left[\frac{4 \alpha_X m_\phi}{M_X v^2_\text{rel}}\right]^2 \lesssim  1\, \text{cm}^2 \times \frac{M_X}{\text{g}} \qquad \alpha_X \equiv N_X^2 \frac{g_\chi^2}{4\pi}
\eeq
where W is the Lambert W function. The inequality again represents the Bullet Cluster constraint. The self-interaction bound for a light mediator is then
\beq\label{eq:sidmlongrange}
g_\chi \lesssim \left(\frac{\pi}{4}\frac{\bar{m}_\chi^4 v_{rel}^4  }{ M_X}\times \frac{1\, \text{cm}^2}{g}\right)^{1/4} e^{\frac{1}{4}\sqrt{\frac{ m^2_\phi M_X}{\pi}\times  \frac{1\,\text{cm}^2}{g}}}.
\eeq
The exponential term indicates that the bound quickly becomes negligible when the range of the mediator becomes small compared to the Bullet Cluster bound on the scattering length. This occurs for
\beq
 M_X \gtrsim 10^{16}\GeV \left(\frac{\eV}{m_\phi}\right)^2.
\label{eq:MomTransSIDM}
\eeq
In this regime the light mediator $\phi$ does not contribute to the SIDM bound, though the constraint in Eq.~\eqref{eq:GeoSIDM} still applies, due to the short range DM-DM force that binds the constituents of the ADM nugget.

The bounds in \Eqs{eq:GeoSIDM}{eq:MomTransSIDM} do not apply if the nuggets make up less than roughly $10\%$ of the dark matter. Given the relative complexity of the dark sector and the non-trivial formation history, this is rather plausible, in particular if there are bottlenecks in the dark fusion processes \cite{Gresham:2017cvl}.  To illustrate the variation in the direct detection phenomenology, we will consider a benchmark with $\mxbar=10$ MeV, which is excluded by self-interaction bounds for $M_X\lesssim 4\times 10^6$ GeV if all of the DM is made up out of a nugget of a single mass. For this benchmark we therefore assume that the ADM nuggets we consider for direct detection compose 1\% of the total DM density. All direct detection limits can be trivially rescaled to the desired subcomponent fraction, including 100\%, in the large $M_X$ part of the parameter space.

 While the SIDM constraints do not apply for subcomponent DM, there are additional consistency conditions on $g_\chi$ which must be satisfied even in a subcomponent scenario: for example, the potential for a light scalar mediator will be modified due to the presence of the background charge density sourced by the nugget constituents.  In particular, one expects the mass of a light field to receive an in-medium correction of the order $\delta m_\phi^2 \sim \frac{g_\chi^2}{\mxbar} \langle \overline \chi \chi\rangle$ with $ \langle \overline \chi \chi\rangle\sim \bar{m}_\chi^{3}$ from the background $\chi$-field in the nugget. If this correction reduces the range of the mediator inside the nugget to be comparable or smaller than the radius of the nugget itself ($\delta m_\phi^{-1} \gtrsim R_X$), the scattering cross section is dramatically altered\footnote{This is akin to the Debye screening mass that a photon develops in a charged plasma.}. To avoid this, we require
  \beq\label{eq:screening}
  g_\chi \lesssim N_X^{-1/3} \, .
  \label{eq:stability}
  \eeq
When the coupling becomes larger than this, the in-medium screening effects will alter the experimental reach, including its dependence on $M_X$, and so we will limit ourselves to regions of parameter space where Eq.~\eqref{eq:screening}  is satisfied. If one instead assumes a repulsive interaction instead of an attractive one, the ADM nugget becomes unbound unless, again, $g_\chi \lesssim N_X^{-1/3}$.

  Finally, in Sec.~\ref{sec:electroncoupling} we will briefly consider the special case of a light, kinetically mixed dark photon acting as the mediator. In this case the interaction is repulsive and it may be difficult for large ADM nuggets to form, due to a Coulomb-like barrier sourced by the dark photon. In order for ADM nuggets to undergo fusion in the early Universe, they must have  enough kinetic energy to overcome this repulsive barrier, i.e.
  \beq
 \frac{ \mu_{X_1 X_2} v^2}{2} \gtrsim g_\chi^2\frac{ N_{X_1}N_{X_2}}{R_{X_1}+ R_{X_2}}
  \eeq
  where $\mu_{X_1 X_2}$ is the reduced mass of the two nuggets and we have allowed for the possibility that fusion occurs between nuggets of different sizes.  Assuming fusion occurs either between nuggets of the same size or between a nugget and a single constituent, the barrier can be overcome if 
  \beq
  g_\chi \lesssim \begin{cases}N_X^{-5/6} \qquad & \text{Two identical nuggets} \\ N_X^{-1/3} \qquad & \text{Constituent and a nugget}
  \label{eq:gchifusions}
  \end{cases}\,\, .
  \eeq
 One might wonder whether instead of overcoming the barrier classically, it would be possible to have fusion occur via quantum mechanical tunneling through the barrier. In this case, the constraint on $g_\chi$ involves calculating the Gamow factor and requiring that the transmission probability is a sizable fraction of unity. The corresponding constraint on the coupling is
   \beq
  g_\chi \lesssim \begin{cases}N_X^{-5/4} \qquad & \text{Two identical nuggets} \\N_X^{-1/2} \qquad & \text{Constituent and a nugget}
  \end{cases}
   \label{eq:gchiGamow}
  \eeq
  where in Eqs.~(\ref{eq:gchifusions})-(\ref{eq:gchiGamow}), we assume that the fusion occurs in a thermal bath whose temperature is on the same order as $\mxbar$. For scattering in this type of environment, fusion due to quantum mechanical penetration of the barrier is negligible until the barrier can be overcome classically.  Therefore, we take Eq.~\eqref{eq:gchifusions} as indirect constraints on $g_\chi$ when considering dark photons. While model specific contributions can change the prefactors in Eq.~\eqref{eq:stability} and Eq.~\eqref{eq:gchifusions}, the scaling as a function of $N_X$ will remain.

\subsection{Direct detection of composite Dark Matter}

For a heavy mediator, the potential seen by a point-like DM particle is simply the sum of the potential for each of the scattering centers in the target. The part of the Hamiltonian governing this contact interaction is
\begin{equation}\label{eq:neutronpotential}
\mathcal{V}(\bfr )=\sum_{J=1}^{N} \mathcal{V}_J(\bfr _J-\bfr )=\frac{2\pi b_{Xt}}{\mu_{Xt}}\sum_{J=1}^{N} \delta(\bfr _J-\bfr )
\end{equation}
or in Fourier space
\begin{equation}
\mathcal{V}(\bfq )= \frac{2\pi b_{Xt} }{\mu_{Xt}}\sum_J^{N} e^{i \bfq \cdot \bfr_J}.
\end{equation}
where $J$ sums over all scattering centers with location $\bfr_J$, $\mu_{Xt}$ is the reduced mass of the DM and the SM target particle,\footnote{Throughout, we use the convection that lowercase letters indicate constituents and capital letters indicate the bound states. E.g.~$m_n$ and $m_N$ refer to the nucleon mass and nucleus mass respectively.} which we take to be a nucleon or an electron; we will generalize to nuclei in the next section. The DM-target scattering length is $b_{Xt}$, again defined as either the DM-nucleon or DM-electron scattering length, depending on the model and process under consideration. For simplicity we have assumed a single type of scattering center in the target, by pulling the $b_{Xt}/\mu_{Xt}$ factor in front of the sum; the generalization to multiple types of scattering centers is straightforward and will be included when appropriate ({\em e.g.}~for a GaAs target). The potential is normalized such that the cross section for a probe on a single, free scattering center is
\beq\label{eq:refcros}
\overline \sigma_0 \equiv 4\pi b_{Xt}^2.
\eeq
In other words, $\overline\sigma_0$ is the cross section for the scattering of a hypothetical DM nugget with radius zero off a SM nucleon or electron. Since this quantity is the analogue to the familiar DM-nucleon/DM-electron cross section for elementary DM, we will therefore use it to parametrize the constraints. In particular for DM that couples to nucleons, $\overline\sigma_0$ is defined as the cross section off a single, free nucleon, to allow for convenient comparison between targets with different atomic mass numbers. 

Given that we will be comparing a variety of different target materials, it is helpful to explicitly separate the DM kinematics from the properties of the target by defining the structure function of the target by
\begin{equation}\label{eq:structurediff}
S(\bfq,\omega)\equiv \frac{1}{N} \sum_{f} \left|\sum_J^{N}\bra \lambda_f |e^{i \bfq\cdot \bfr _J}|\lambda_i \ket\right|^2 \delta (E_{f}-E_i-\omega)
\end{equation}
which is normalized to be an intrinsic quantity. The $\bra\lambda_{i,f}|$ are the initial and final states of the target, and the $E_{i,f}$ their corresponding energies; $\bfq$ and $\omega$ are respectively the momentum transfer and the energy deposited by the DM. The states $\bra\lambda_{i,f}|$, and therefore the function $S(q,\omega)$, are highly material dependent and must be computed on a case-by-case basis. In general, $S(\bfq,\omega)$ depends on the direction of $\bfq$, in particular for long-wavelength scattering in anisotropic crystals. However in this paper we will always work in the isotropic approximation, where $S(\bfq,\omega)$ only depends on $q\equiv|\bfq|$ and $\omega$. It is worth noting that this prescription is correct only as long as the Born approximation is valid, i.e. as long as the total cross section remains smaller than the geometric cross section. This condition is violated in part of the parameter space, as we will comment on when appropriate.

By Fermi's golden rule, the differential cross section of a point-like probe with the target is then
\begin{equation}\label{eq:pointlike}
\frac{d\overline\sigma}{dq\,d\omega} = \overline \sigma_0 \frac{q}{2v^2\mu_{Xt}^2} \,  |F_{\text{med}}(q)|^2\,S(q,\omega),
\end{equation}
where $v$ is the DM velocity in the lab frame. We have included the form factor induced by the mediator particle, which we take to be
\begin{equation}\label{eq:medformfactor}
F_{\text{med}}(q) =\left\{ \begin{array}{ll} 
1& \qquad \text{(heavy mediator)}\\
 q_0^2/q^2 & \qquad \text{(light mediator)}
 \end{array}\right. ,
\end{equation}
where $q_0$ is a reference momentum transfer.  The typical momentum transfer is often proportional to $\mu_{Xt}$, which depends on the target. We will not study the intermediate regime, where the mediator being heavy or light varies from experiment to experiment --   we will assume that the mediator mass is such that either the heavy or light case applies for all experiments shown in a given plot. In terms of the underlying model parameters $\overline \sigma_0$ is given by
\begin{equation}\label{eq:modeleq}
\overline \sigma_0= N_X^2 \frac{g_t^2 g_\chi^2}{4\pi}\times \left\{ \begin{array}{ll} \mu_{Xt}^2/m_\phi^4\qquad\text{(heavy mediator)}\\
\mu_{Xt}^2/q_0^4\qquad\;\,\text{(light mediator)}\end{array}\right.
\end{equation}  
with $g_t$ standing for $g_e$ or $g_n$, depending on the case at hand. The $N_X^2$ factor makes the coherence over the nugget constituents explicit. Finally, with Eq.~\eqref{eq:modeleq} it becomes manifest that the physical cross section for a light mediator in Eq.~\eqref{eq:pointlike} is independent of the arbitrary parameter $q_0$.
\begin{figure}[b] 
\begin{center}
\includegraphics[width=0.5\textwidth]{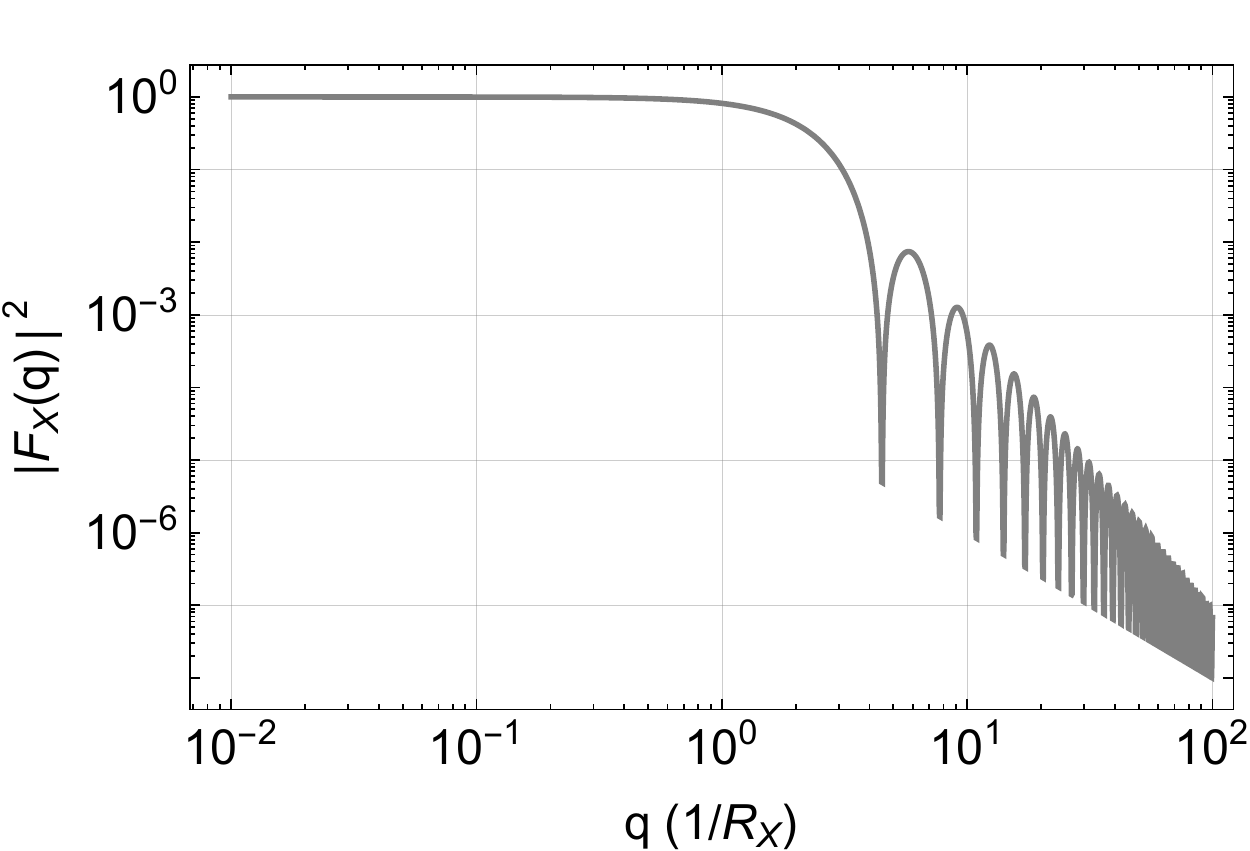}
\caption{Nugget form factor, with $q$ in units of $1/R_X$.  }
\label{fig:formfactorplot}
\end{center}
\end{figure}

Moving on to the scattering of extended nuggets, we must introduce a form factor to account for the non-zero radius of the DM particle, as laid out in \cite{Hardy:2015boa}. For elastic scattering with uniform couplings to all constituents, the form factor is given by the Fourier transform of the mass density, $\rho_X$,
\beq
F_X(\textbf{q})=\frac{1}{M_X}\int\! d^3\textbf{r}\; e^{i\textbf{q}\cdot\textbf{r}} \rho_X(\textbf{r}).
\eeq
Further assuming that the constituents of nuggets are uniformly distributed, the mass density is simply a three dimensional spherical top hat function and thus the form factor is
\beq \label{eq:dmformfactor}
F_X(q)=\frac{3j_1(qR_X)}{qR_X}=\frac{3(\sin(qR_X)-qR_X\cos(qR_X))}{(qR_X)^3}
\eeq
where $j_1(x)$ is a spherical Bessel function of the first kind (see Fig.\ref{fig:formfactorplot}). The differential scattering cross section of an ADM nugget can then be written as
\begin{equation}\label{eq:nuggetdiff}
\frac{d\sigma}{dq\,d\omega} = \frac{d\overline\sigma}{dq\,d\omega} |F_X(q)|^2
\end{equation}
where $d\overline \sigma/dq\,d\omega$ is the differential cross section of a point-like dark matter particle with the same mass and coupling strength as the ADM nugget, as defined in Eq.~\eqref{eq:pointlike}. We have normalized the form factor such that $F_X(0)=1$, such that we manifestly recover the point-like limit if $qR_X\ll1$. In this regime all constituents contribute coherently and the scattering rate scales with the square of the number of constituents, as is evident from Eq.~\eqref{eq:modeleq}. If the inverse momentum transfer is smaller than the radius but still larger than the interparticle spacing of the nugget, coherence is lost and the form factor drops rapidly. Finally, once the momentum transfer is larger than the inverse interparticle spacing, we enter the regime of deep inelastic scattering (DIS), and the form factor description breaks down.  Given the relatively low speed of the DM in the galaxy, the maximum momentum transfer in a single collision is $\OO( m_{N}\, v)\sim100$ MeV, such that DIS is only relevant for $\mxbar\ll100$ MeV. If the interior of ADM nugget is resolved in the collisions, the rate moreover is always dominated by the lowest accessible momentum transfer, such that the DIS contributions are a negligible correction. The one exception occurs for high threshold, noble liquid experiments when $\mxbar \lesssim 10$ MeV; here, the experimental threshold pushes the scattering into the crossover region between the form factor and DIS descriptions. When relevant, we will comment on this effect when presenting our results. 

We can combine Eq.~\eqref{eq:structurediff} and Eq.~\eqref{eq:nuggetdiff} to obtain the differential rate per unit of exposure, as seen by the experiments
\begin{equation}\label{eq:masterform}
\frac{dR}{d\omega}= \frac{\rho_X}{M_X}\frac{n_T}{\rho_T}\frac{\overline \sigma_0}{2\mu_{Xt}^2}\int\! dv\; \frac{1}{v} f(v) \int\! d q\; q\;|F_X(q)|^2\;|F_{\text{med}}(q)|^2\; S(q,\omega)
\end{equation}
with $\rho_X\approx 0.3\ \text{GeV}/\text{cm}^3$ the DM mass density in the galaxy, $\rho_T$ the mass density of the target material, and $n_T$ the number of available scattering centers in the target.  Here, $f(v)$ is the velocity distribution of the DM in the frame of the earth, given by
\begin{align}
	f(v) &= \frac{1}{N_0} \frac{\pi v v_0^2}{v_e} \times \left\{\begin{array}{ll}
	 e^{-(v+v_e)^2/v_0^2}\;\left(e^{4v\, v_e /v_0^2}-1\right)	& v <v_{esc}-v_e\\
	 e^{-(v-v_e)^2/v_0^2}-e^{v_{esc}^2 /v_0^2}		& v_{esc}-v_e<v <v_{esc}+v_e\\
	 0										& v >v_{esc}+v_e
	\end{array}\right. ,
	 \label{eq:velodist} \\
	&N_0 \equiv \pi^{3/2} v_0^3 \left[ {\rm erf} ( \tfrac{v_{\text{esc}}}{v_0} ) - \tfrac{2}{\sqrt{\pi}}\tfrac{v_{\text{esc}}}{v_0} \exp\left( -( \tfrac{v_{\text{esc}}}{v_0} )^2 \right) \right] ,
\end{align}
with $v_0=220$ km/s, $v_e=240$ km/s, truncated by the escape velocity $v_{\text{esc}}=550$ km/s. 

 Energy and momentum conservation enforce target-dependent boundary conditions on the $q$-integral, depending on whether the DM scatters with a nucleus, a phonon or a bound electron. Finally, different detectors are sensitive to different energy ranges, which in turn restricts the integration range over $\omega$. We will elaborate on all these differences and their impact in the following sections, but it should be already clear from Fig.~\ref{fig:formfactorplot} that in particular the boundary conditions on $|\bfq|$ will have a huge impact on the rate: if the threshold of a particular experiment is too high to allow for coherent scattering, its reach will be greatly reduced.

\section{Nucleon couplings\label{sec:nucleoncoupling}}
In this section we consider the case where the nuggets have a coupling to SM nucleons only.  The nucleon couplings give rise to a nuclear recoil, or in the case that the momentum transfer is smaller than the inter-atom spacing, coherent phonon excitations.  As was shown in Refs.~\cite{Gresham:2017zqi, Gresham:2017cvl, Gresham:2018anj}, there is enormous variation in dark nugget formation histories and resulting size. For this reason, it does not seem warranted at this point to carefully recast all existing limits on elementary dark matter to the composite case. Instead we consider a handful of simplified experimental concepts which are representative of the larger set of experiments and ideas for (future) dark matter direct detection, as listed in Table~\ref{tab:nucleonexperiments}.

 In particular, we compare high threshold, high exposure noble liquid experiments with low threshold, low exposure experiments such as semi-conductor or superfluid helium targets.  (For a recent overview, see \cite{Battaglieri:2017aum}.)   For nuggets on the higher end of the mass range, as we will see, only the low threshold experiments benefit from the coherent enhancement over nucleon number, such that they can be competitive despite their much lower exposure. This implies a high degree of complementarity between both approaches, which we quantify below. The calculations are qualitatively different depending on whether the scattering takes place in the nuclear recoil regime, or in the regime where direct phonon production dominates and so we treat both cases separately.

\newcolumntype{C}[1]{>{\centering\let\newline\\\arraybackslash\hspace{0pt}}m{#1}}
\begin{table}[t]

\begin{tabular}{p{4cm}C{4cm}C{3cm}C{3cm}C{2cm}}\hline
Experiment&Exposure (kg-year)&Threshold&Timescale&$N_{\mathrm{events}}$\\\hline
XENON1T \cite{Aprile:2017iyp}&98&5 keV&existing limit&3\\
CDMSlite \cite{Agnese:2017jvy}&0.2&0.5 keV&existing limit&BD\\\hline
Liquid Xe&$1.5\times 10^4$& 5 keV &in progress&10\\
semi-conductor (Ge)&100&50 eV&in progress&$10^4$\\
semi-conductor (Si) &1&10 meV&R\&D needed&  3\\
superfluid He (recoil) &1&1 meV&R\&D needed&  3\\
superfluid He (phonon)&1&1 meV&R\&D needed&  3\\
GaAs (phonon) &1&1 meV&R\&D needed&  3\\
\hline
\end{tabular}
\caption{Overview of assumed exposure and thresholds for the existing and future experiments considered in this work, for DM-nucleon couplings. $N_{\mathrm{events}}$ refers to the number of expected signal events that were assumed to estimate the 90\% exclusion limits, where ``BD'' refers to a bin dependent analysis. (See text for details.)
\label{tab:nucleonexperiments} }
\end{table}

\subsection{Nuclear recoils} \label{nuclearrecoilssection} 
For ordinary nuclear recoils, the structure function in Eq.~\eqref{eq:masterform} is
\begin{equation}\label{eq:nucleonxsec}
S(q,\omega)= A^2 |F_N(q)|^2 \delta\left(\omega-\frac{q^2}{2m_N}\right),   
\end{equation}
with  $A$ the atomic mass number of the nucleus and $m_N$ the mass of the nucleus. $F_N(q)$ is the Helm form factor of the nucleus 
\beq
F_N(q)\equiv\frac{3j_1(qR_N)}{qR_N}e^{-q^2s^2/2},
\eeq
with skin depth $s\simeq$ 0.9 fm and $R_N\approx A^{1/3}\times 1.2$ fm. We have assumed that the scattering length is the same for protons and neutrons.

 We take XENON1T \cite{Aprile:2017iyp} as a representative for the existing limits from the large exposure experiments;\footnote{The kinematic restriction in Eq.~\eqref{eq:nucleonxsec} is lifted if the nucleus emits a photon during its recoil \cite{Kouvaris:2016afs}, or by emitting a Migdal electron \cite{Ibe:2017yqa}. For light DM these processes add sensitivity for detectors of which the threshold would normally be too high to detect the regular nuclear recoil, as is evident from a recent LUX analysis \cite{Akerib:2018hck}. For ADM nuggets, one can verify that the bremsstrahlung process never provides more reach than nuclear recoils, regardless of $R_X$. The recast of the Migdal effect is more involved, and we leave this for future studies. } PandaX-II~\cite{Cui:2017nnn} and LUX~\cite{Akerib:2016vxi} have set very similar bounds. XENON1T reaches its maximal signal efficiency for $10\,\mathrm{keV}\lesssim E\lesssim$ 45 keV, but is partially efficient for substantially lower/higher recoil energies. To model this effect, we use the full efficiency curve supplied in \cite{Aprile:2017iyp} in our calculations. For our approximate reinterpretation, we use the ``reference region'' defined in \cite{Aprile:2017iyp}, which contains roughly 50\% of the signal and an expected background of 0.36 events. No events were observed and so we take 3 signal events in this region as our approximate limit. We consider \mbox{CDMSlite \cite{Agnese:2017jvy}} as an example of an existing, low-threshold experiment with a semi-conductor target. (Other experiments with competitive results on low-threshold nuclear recoils are CRESST-II~\cite{Angloher:2015ewa} and NEWS-G~\cite{Arnaud:2017bjh}.) CDMSlite had an exposure of 70 kg-days in their second run, with a threshold of roughly 0.5 keV and a maximal signal efficiency leveling around 50\%. We use the full efficiency curve provided by the collaboration, which we unfold to nuclear recoil energies using the Lindhard formula with the parameters given in \cite{Agnese:2017jvy}. The analysis had a large number of background events, divided over four non-overlapping energy bins. For each bin, we extract a limit on the cross section assuming that all background events can be interpreted as signal, and select the strongest such limit for each mass point. This procedure reproduces the CDMSlite limit to within 50\% for DM masses above roughly 10 GeV. For the part of the phase space near the detector threshold the agreement is worse, however our estimate is always conservative.

Loosely modeled after LUX-ZEPLIN \cite{Akerib:2018lyp}, we estimate the reach for future high exposure experiments by assuming a 15 ton-year exposure and a 5 keV (75 keV) lower (upper) energy threshold. We further assume a background expectation of 5 events in the signal region, which corresponds to an expected 90\% exclusion of roughly 10 signal events if no background subtraction is attempted. For a future, low threshold semi-conductor experiment we consider a Ge target with 100 kg-year exposure, a threshold of 50 eV and $10^4$ expected background events. These parameters are loosely modeled after the superCDMS \cite{Agnese:2016cpb} projections. 

Lastly, we consider a few ultra-low threshold options, for which the R\&D is still ongoing: A miniaturized silicon detector with transition edge sensor (TES) read-out, for which we take the dynamic range between 10 meV and 1 eV, and a long-term exposure of 1 kg-year, which may eventually be achieved by multiplexing the detector. In the same vein, we include a superfluid helium target in nuclear recoil mode \cite{Guo:2013dt,Schutz:2016tid,Hertel:2018aal} with a dynamic range between 1 meV and 100 meV; superfluid helium in the phonon mode will be discussed in the next section. For both concepts, we (optimistically) assume negligible irreducible background. At least for radiogenic backgrounds, this is expected to be true due to the suppressed background spectrum at low recoil energies (see e.g.~\cite{Hochberg:2015fth}) and the relatively low exposure. It is moreover straightforward to approximately rescale our projections for different background assumptions. Other future proposals, which we do not discuss here, include searches for anomalous color centers~\cite{Essig:2016crl,Budnik:2017sbu}, magnetic bubble chambers \cite{Bunting:2017net} and excitations in molecular gases \cite{moleculargass}.

In the remainder of this section we will present a number of analytic, approximate formulae for the rate.  This will allow us to understand the features present in the sensitivity curves and to make explicit where low threshold detectors can perform better than the standard noble liquid experiments. We make the following approximations: (i) we set the DM velocity distribution to a delta function centered at $v_0$, (ii) drop the form-factor for the SM nuclei ($F_N\approx1$) and (iii) take $M_X\gg m_N$, such that the reduced mass $\mu_{XN}\approx m_N$. The ADM nugget form factor can further be expanded as
\beq
|F_X(q R_X)|^2 = \begin{cases} 1 + \OO (q^2 R_X^2) &\qquad q R_X \ll 1\\\frac{9}{2 \left(q R_X\right)^4} + \OO\left(\frac{1}{q^6 R_X^6}\right) &\qquad q R_X \gg 1\end{cases}\,\,,
\label{eq:FFExpansion}
\eeq
where the expansion at large momentum transfer only holds when the form factor is integrated against a smoothly varying function of $q$. 

\begin{figure}[t] 
\begin{center} 
\centering
    \begin{minipage}{0.49\textwidth}
        \centering
        \includegraphics[width=1\textwidth]{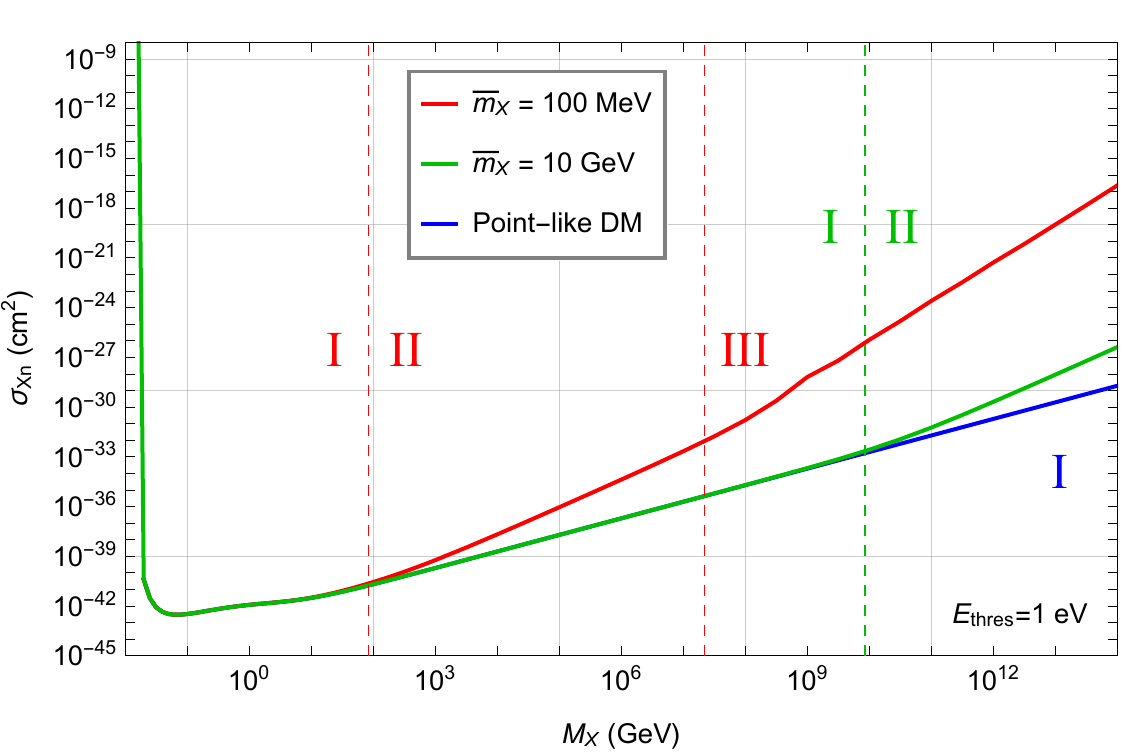} 
    \end{minipage}\hfill
    \begin{minipage}{0.49\textwidth}
        \centering
        \includegraphics[width=1\textwidth]{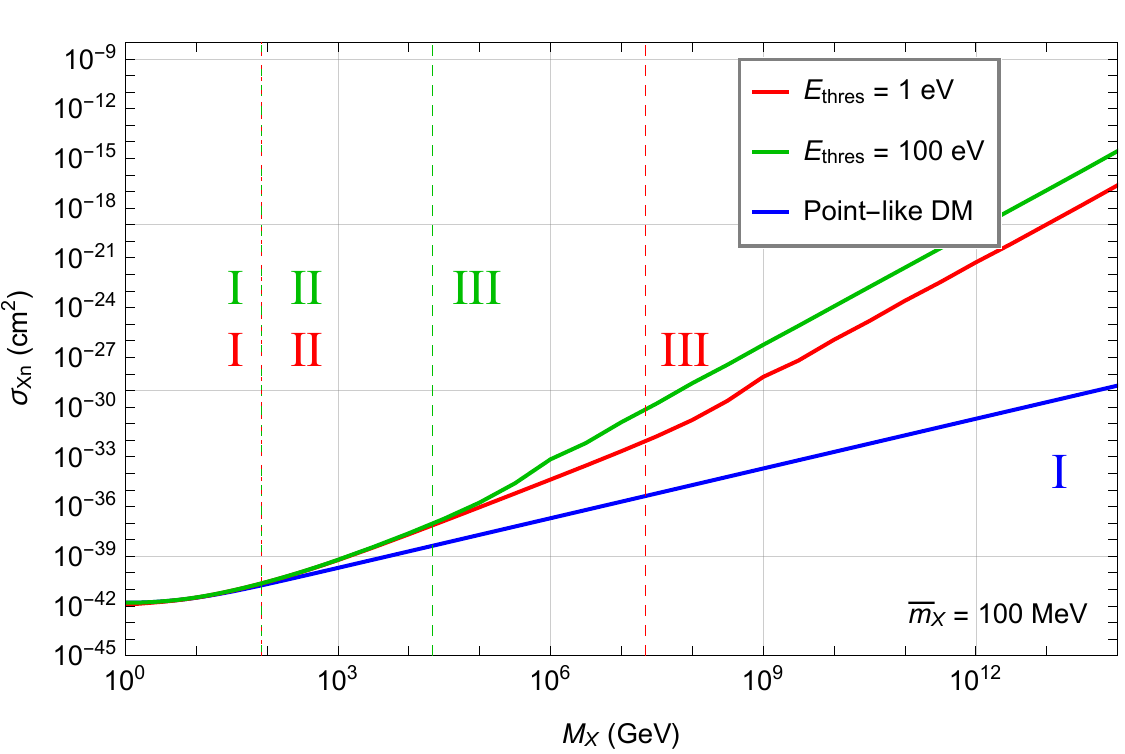} 
    \end{minipage}

\caption{Example reach curves for nuggets and point-like DM particles for He target. Left: Fixed $E_{\text{thres}}=1 $ eV, varying $\mxbar$. Right: Fixed $\mxbar=100$ MeV, varying $E_{\text{thres}}$. Roman numerals indicate the regions where the impact of the form factor differs qualitatively, as described in the text.} \label{fig:hetwokinks}
\end{center}
\end{figure}

With the assumptions outlined above, the maximum and minimum momentum transfer for nuclear scattering is
\beq
q_\text{max} =2 m_N v_0 \mbox{   and   }    q_\text{min} = \sqrt{2 m_N E_\text{thres}},
\eeq
where $E_\text{thres}$ is the experimental threshold.  There are then three regions of the sensitivity curve that each have a different dependence on the ADM nugget mass, as shown in Fig.~\ref{fig:hetwokinks}.  They are:

\begin{itemize}
\item Region I: No form factor suppression,  $q_\text{max}R_X \lesssim 1$.

The ADM nugget radius is always less than the inverse momentum transfer. The SM nucleus therefore cannot probe any of the finite size properties of the ADM nugget and the nugget behaves like a point-like particle for all available phase space, regardless of the threshold of the detector. For heavy DM, this property holds when
\beq\label{eq:regionItoII}
\frac{\mxbar}{18 \pi} \left(\frac{\mxbar}{m_N v_0}\right)^3 \gtrsim M_X.
\eeq

\item Region II: Partial form factor suppression, $q_\text{max}R_X \gtrsim 1 \gtrsim q_\text{min}R_X$.

The ADM nugget radius is larger than the inverse momentum transfer for some of the accessible phase space, and the coherent enhancement only occurs for a fraction of the scattering phase space. For heavy DM, this partial form factor suppression occurs when
\beq
\frac{\sqrt{2}\mxbar}{9 \pi}\left(\frac{\bar{m}_\chi^2}{m_N E_\text{thres}}\right)^{3/2} \gtrsim M_X \gtrsim \, \frac{\mxbar}{18 \pi} \left(\frac{\mxbar}{m_N v_0}\right)^3.
\eeq

\item Region III: Complete form factor suppression, $q_\text{min}R_X \gtrsim 1$.

The ADM nugget radius is larger than the inverse momentum transfer even for the smallest detectable momentum transfer, and thus the SM nucleus can probe the finite size properties of the ADM nugget for all of the accessible phase space. This implies that the coherent enhancement never occurs. For heavy DM, this complete form factor suppression occurs when
\beq
M_X \gtrsim \, \frac{\sqrt{2}\mxbar}{9 \pi}\left(\frac{\bar{m}_\chi^2}{m_N E_\text{thres}}\right)^{3/2}.  \eeq
\end{itemize}

Due to the simplicity of dynamic structure function for nuclear recoil, the detector-specific rate can be well approximated by an analytic expression. As we seek to determine which detectors work best for various regions of the ADM nugget parameter space, we pull out the dependence of the rate on the atomic mass number  $A$ by substituting $m_N=A\, m_n$.

 For a heavy mediator, the fiducial cross section for an ADM nugget with mass $M_X$ and radius $R_X$, scattering off of a target nucleus is
\beq \label{eq:shortrangeapprox}
\sigma^{\text{heavy}}_{NR}&\approx& \overline \sigma_0 \times \begin{cases}\frac{A^2q^2_{\text{max}}}{4v_0^2 m_n^2}=A^4&\qquad \text{Region I} \\
\frac{9A^2}{8R_X^2 v_0^2 m_n^2}&\qquad \text{Region II}\\
\frac{	9 A^2}{8q^2_{\text{min}}R_X^4 v_0^2 m_n^2}=\frac{9 A}{16E_{\text{thres}}R_X^4 v_0^2 m_n^3}&\qquad \text{Region III}\end{cases},
\label{eq:NRMassive}
\eeq
where $\overline \sigma_0$ is the reference cross section for scattering off a single, free nucleon, as defined in Eq.~\eqref{eq:refcros}. In all regions, the cross section receives an $A^2$ enhancement from the coherence over the SM nucleus. In region I, this is supplemented with an $A^2$ phase space enhancement for heavy target nuclei, while for region III the $1/{q^2_{\text{min}}}$ phase space suppression partially cancels the effect of the coherent enhancement over the SM nucleus. 

The effect of the form factor on the reach of two hypothetical experiments with different thresholds is shown in Fig.~(\ref{fig:hetwokinks}), where we labeled the transitions between regions I,~II and III. In both panels, the crossover between regions I and II indicates where the form factor starts to reduce the reach relative to the point-like DM case. The left hand panel shows that this transition occurs sooner for lower values of $\mxbar$, which correspond to a larger nugget. In the right hand panel we see that the transition between regions I and II is independent of $E_{\mathrm{thres}}$, as expected from Eq.~\eqref{eq:regionItoII}.

The dependence of the cross section on the dark nugget radius in the three different regions can be explained by simple arguments. In region I, where the nugget behaves like a point particle, there is no dependence on $R_X$. In region III, where the form factor is nontrivial for the entirety of the phase space, the expansion in Eq.~\eqref{eq:FFExpansion} dictates that the cross section will scale as $1/R_X^4$. Lastly, in the intermediate region II, where the form factor only affects high momentum transfer scattering, the cross section scales with $1/R_X^2$. This is because the largest momentum transfer that is unaffected by the form factor is roughly $1/R_X$, such that one can obtain an estimate for the cross section by substituting $q_\text{max} \rightarrow 1/R_X$ in the expression for region I.

For interactions between the ADM nuggets and SM nuclei via a light mediator,  these conclusions change qualitatively: the additional form factor in Eq.~\eqref{eq:medformfactor} further weights the matrix element towards lower momentum transfer and favors low threshold experiments over their higher threshold counterparts.  The approximate fiducial cross section in this case is
\beq \label{eq:longrangeapprox}
\sigma^{\text{light}}_{NR}&\approx& \overline \sigma_0 \times q_0^4\times \begin{cases}
\frac{A^2 }{4q_\text{min}^2 v_0^2 m_n^2}=\frac{A }{8E_\text{thres} v_0^2 m_n^3}& \text{Region I \& II} \\
\frac{3A^2}{8R_X^4 v_0^2 m_n^2 q_{\text{min}}^6}=\frac{3}{64 AR_X^4 v_0^2 m_n^5 E_{\text{thres}}^3}& \text{Region III}\end{cases},
\eeq
where $q_0=v_0 m_n$ is an arbitrary reference momentum, to render the definition of $\overline \sigma_0$ in Eq.~\eqref{eq:modeleq} IR finite.  The cross section is even more biased towards the low momentum transfer regime, as is evident from the $q_{\text{min}}$ factors in the denominator of both expressions. This implies that to leading order, there is no difference between regions I and II, since in both cases the process is still coherent near the low momentum threshold. The phase space distribution also changes the dependence on the atomic mass number $A$, making heavy nuclei detectors less favorable, as compared to the heavy mediator case in Eq.~\eqref{eq:shortrangeapprox}.  For both Region I and Region II, where the dominant contribution to the scattering rate is unaffected by the form factor, there is no dependence on $R_X$. It is only in Region III, where the form factor affects the entirety of the phase space, do we see a $1/R_X^4$ dependence that arises from the expansion of Eq.~\eqref{eq:FFExpansion}.

\subsection{Collective Modes}\label{sec:coherentmodes}

When the inverse momentum transfer is larger than the inter-particle spacing, the response of the target is qualitatively different than what was considered in the previous section: the target can no longer be treated as a collection of non-interacting nuclei and instead one has to integrate out the atoms and use an effective description of a phonon gas at zero temperature. While scattering at such low momentum occurs for all types of dark matter, this contribution is typically heavily phase space suppressed for heavy dark matter. Large dark nuggets are an exception to this rule, as the coherent $N_X^2$ enhancement to the cross section at low momenta can overcome the phase space suppression in certain regions of the parameter space. A similar effect can occur when the scattering is due to a light mediator. Calculations of the single phonon production have been performed for the polar materials GaAs and sapphire \cite{Knapen:2017ekk,Griffin:2018bjn} and for two phonon/roton production in superfluid helium  \cite{Schutz:2016tid,Knapen:2016cue}. 

For superfluid helium, we estimate the reach with the analytic expressions for the dynamic structure function in Refs.~\cite{Schutz:2016tid,Knapen:2016cue} which match the simulation data \cite{Krotscheck2015} to within an order of magnitude. In particular, in the region where $|\bfq| \lesssim 1/\text{\AA}$, the structure function can be approximated as 
\beq
S(q,\omega) \approx \frac{q}{2 m_{\rm He} c_s} \delta(\omega-\omega_q) +  \frac{7 m_{\rm He}^{5/2}}{60 \pi^2 \rho_0}\frac{c_s^4 q^4}{\omega^{7/2}}
\eeq
where the first and second terms represent single and double phonon production respectively. However,  for $q \lesssim$ keV and optimistic energy thresholds, the single phonon component is always too soft to contribute to the detection rate, as its energy is $\omega_q\approx c_s q$ where $c_s\approx 10^{-6}$ is the speed of sound in superfluid He. We therefore rely on double phonon production to detect ADM nuggets in superfluid helium when the momentum transfer is below the nuclear recoil threshold. 

For scattering off phonon excitations in a crystal, the structure function is 
\beq
S(\textbf{q},\omega)=\frac{1}{2}\sum_\nu \frac{|F_\nu(\textbf{q})|^2}{\omega_{\nu,\textbf{q}}}\delta(\omega_{\nu,\textbf{q}}-\omega)
\eeq
with $F_\nu(\textbf{q})$ and $\omega_{\nu,\textbf{q}}$ are respectively the phonon form factor and energy corresponding to the phonon branch $\nu$ and momentum $\bfq$. We consider the example of GaAs, which has a cubic crystal structure with one gallium and one arsenic atom in its primitive cell, which give rise to 6 independent phonon modes. To leading order, only the longitudinal modes contribute with the form factor
\beq\label{eq:phononformfactor}
|F_\pm(\textbf{q})|^2 \approx \frac{q^2}{2} \left \lvert \frac{A_{\textrm{Ga}}}{\sqrt{m_{Ga}}} e^{i\textbf{r}_{\textrm{Ga}} \cdot \textbf{q}} \pm \frac{A_{\textrm{As}}}{\sqrt{m_{As}}} e^{i\textbf{r}_{\textrm{As}} \cdot \textbf{q}} \right \rvert^2
\eeq
where the $+$ and $-$ indicate the longitudinal acoustic (LA) and longitudinal optical (LO) mode, respectively. $A_{Ga}$ and $A_{As}$ are the atomic mass numbers for gallium and arsenic, while $\bfr_{Ga}$ and $\bfr_{As}$ are the locations of the gallium and arsenic atom with respect to the origin of the primitive cell. For a rough estimate, one can take $\textbf{r}_{\textrm{Ga,As}} \cdot \textbf{q}\approx0$. The LO mode has energy $\omega_{\text{LO}}\approx 30$ meV, which is to good approximation momentum independent. For $q\lesssim 1$ keV, the LA mode has a linear dispersion relation $\omega_{\text{LA}}=c_s q$ with the sound speed $c_s\approx2\times 10^{-5}$. We refer to \cite{Griffin:2018bjn} for details on the derivation of Eq.~\eqref{eq:phononformfactor}. 
 
Identical to the nuclear recoil case in the previous section, there are three qualitatively different regimes in the parameter space, set by the hierarchy between $q_{\text{min}}$, $q_{\text{max}}$ and $1/R_X$. The main difference is that for phonon modes $q_{\text{min}}$ can be substantially lower than for nuclear recoil, thus enlarging the region of parameter space where coherence over the ADM nugget is possible (region I). We illustrate this for the case of the LA phonon in GaAs with simplified expressions for the fiducial cross section. The analogous expressions for the LO mode and helium can be easily derived. For the LA mode,
\beq
q_{\text{min}}=E_{\text{thres}}/c_s, \qquad q_{\text{max}}=q_c
\eeq
with $q_c\approx 1$ keV the momentum transfer at which the single phonon description starts to break down. For $q\gtrsim q_c$ one has to match to the nuclear recoil regime by including higher order corrections to the matrix element, which we do not attempt here. For a heavy mediator, the fiducial cross section is
\beq
\sigma^\text{heavy}_{LA}&\approx& \overline \sigma_0\times\frac{A}{3 m_n^3  v_0^2 c_s}\begin{cases}q_c^3&\qquad \text{Region I}  \\
\frac{9\pi}{2 R_X^3}&\qquad \text{Region II}\\
\frac{27c_s}{2 R_X^4E_\text{thres}} &\qquad\text{Region III} \end{cases}\,\, .
\eeq
where we use that $A\approx A_{\text{Ga}}\approx A_{\text{As}}$. Due to the strong $q_c^3/m_n^3$ phase space suppression in the coherent regime (region I), we find that phonons in GaAs are never competitive with nuclear recoils in superfluid helium for a heavy mediator. The same conclusion holds for phonons in superfluid helium. For a light mediator on the other hand, the fiducial cross section is
\beq
\sigma^\text{light}_{LA}&\approx& \overline \sigma_0\times q_0^4\times \frac{A}{  m_n^3v_0^2 c_s}\begin{cases}\frac{c_s}{E_\text{thres}} &\text{Region I \& II}  \\
\frac{9c_s^5}{10 R_X^4E_\text{thres}^5}&\text{Region III} \end{cases}
\eeq
This scaling is much more favorable for a low threshold, phonon based detection principle and, as we shall see in the next section this detection strategy can be competitive with high threshold, high exposure experiments.

\subsection{Results\label{sec:nucleonresults}}

We now turn to a quantitative comparison of the numerical results. As discussed extensively in the previous sections, the key quantity is the radius of the ADM nugget, given by
\beq
R_X = \ \left(\frac{9\pi}{4}\frac{M_X}{\bar{m}_\chi^4}\right)^{1/3}.
\label{eq:radiusformula2}
\eeq
In our reach estimate, we fix $\mxbar$ to a few benchmark values and vary the ADM nugget mass $M_X$. For fixed $\mxbar$, the radius therefore grows with increasing $M_X$, while for fixed $M_X$, smaller $\mxbar$ correspond to a larger, less dense ADM nugget. We illustrate the direct detection phenomenology with the limiting cases of $\mxbar=$ 10 MeV and $\mxbar=$ 10 GeV; the  $\mxbar =$ 10 MeV benchmark roughly corresponds to the lowest value for the constituent mass for which one can imagine a plausible formation history, consistent with Big Bang Nucleosynthesis. As discussed in Sec.~\ref{sec:otherconstraints}, this benchmark is also in tension with self-interaction bounds if $M_X\lesssim 4\times 10^6$ GeV, which leads us to assume that the ADM nuggets make up 1\% of the total DM mass density.  This bound does not apply for the $\mxbar=$10 GeV benchmark, which will roughly correspond to the highest value for which low threshold experiments can add sensitivity, for a heavy mediator.

\begin{figure}[p]
\begin{center}
\includegraphics[width=0.75\textwidth]{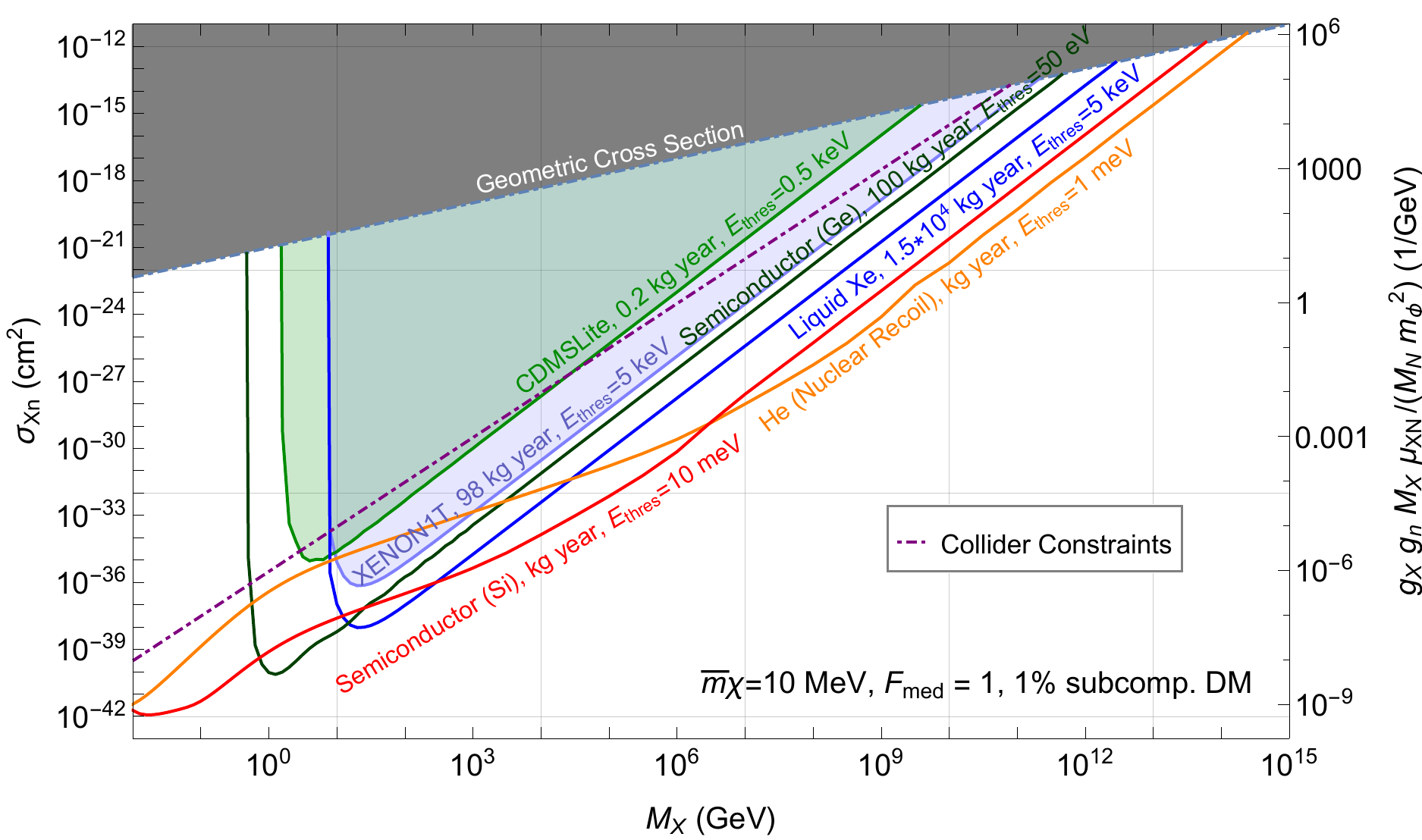}
\caption{Existing and projected reach of experiments in Table~\ref{tab:nucleonexperiments} for heavy mediator coupled to nucleons and nuggets with 10 MeV constituents. We assume nuggets are 1\% of the DM density due to SIDM constraints; the dot-dashed purple curve indicates the LHC constraint on the UV completion in \eqref{eq:colliderbound}.  }
\label{fig:nucleoncoupling_constituent1MeV}
\end{center}
\end{figure}

\begin{figure} [p]
\begin{center}
\includegraphics[width=0.75\textwidth]{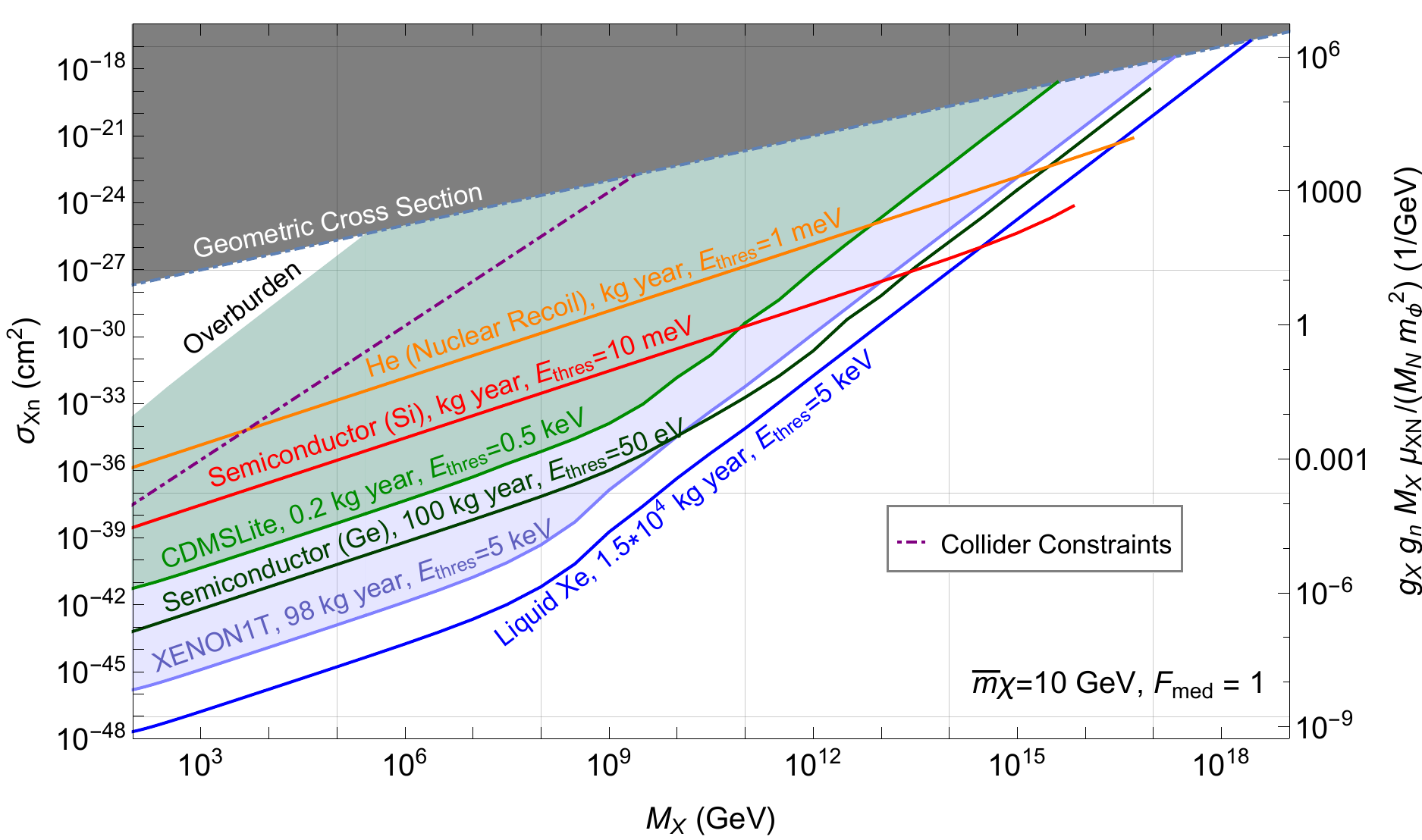}
\caption{Existing and projected reach for the experiments described in Table~\ref{tab:nucleonexperiments},  for heavy mediator interacting with nucleons and nuggets with 10 GeV constituents. The white triangle on the left marks where rock overburden eliminates underground direct detection sensitivity. Constraint curves terminate when the expected number of ADM nuggets passing through the detector drops below one per year.}
\label{fig:nucleoncoupling_constituent10GeV}
\end{center}
\end{figure}

The reach for a heavy mediator is shown in Figs.~\ref{fig:nucleoncoupling_constituent1MeV} and \ref{fig:nucleoncoupling_constituent10GeV} for nuggets with constituent masses $\mxbar=$ 10 MeV and $\mxbar=$ 10 GeV respectively. In the gray region the cross section exceeds the geometric cross section for an ADM nugget with radius $R_X$; the parameter space there is unphysical. Current constraints are shown by shaded regions. There are parts of the parameter space, specifically at large cross sections and low nugget masses, where no experiment has any reach, as the nuggets are stopped by an overburden equivalent of around $2$ km of rock. This overburden effect depends on $M_X$ due to the ADM nugget's form factor, as computed in Appendix~\ref{app:overburden}. In other parts of the parameter space, multiple interactions in the detector can occur in the same event, which requires a dedicated analysis \cite{Bramante:2018qbc}. We also indicate, in dot-dashed purple, the LHC constraints on the UV completion described in Sec.~\ref{sec:MedInteractionSM}. Lastly, note that the maximum nugget mass that a superfluid helium detector can probe with a kg-year exposure is slightly higher than that which can be probed by a silicon detector with the same exposures because of the difference in the detector volumes.

For $\mxbar=10$ MeV (Fig.~\ref{fig:nucleoncoupling_constituent1MeV}), the ADM nugget is large enough such that the fully coherent regime (Region I) is never available, regardless of the detector. For high threshold, high exposure experiments, even partial coherence is not possible, regardless of $M_X$. The fiducial cross section is therefore heavily suppressed as the scattering is always in Region III.  For a low threshold detector on the other hand, $ q_\text{min}R_X<1$ as long as $M_X\lesssim 10^5$ (Si) or $M_X\lesssim 6\times10^7$ GeV (He), such that coherent scattering remains possible in part of the phase space (Region II). For larger $M_X$ the low threshold detectors also transition to the fully incoherent regime (Region III), as is evident from the change in slope of the Si and He curves in Fig.~\ref{fig:nucleoncoupling_constituent1MeV}. Beyond this point, the form factor suppression is present for all experiments but it is more severe for the high threshold detectors.\footnote{For the liquid xenon detectors the minimum momentum transfer is roughly $\sim$ 10 MeV. For the $\mxbar=10$ MeV benchmark this means that strictly speaking the scattering always occurs in the cross over regime between the form factor description and the DIS description. We expect that the true constraints for the xenon experiments could therefore be somewhat weaker than what is shown in Fig.~\ref{fig:nucleoncoupling_constituent1MeV}. } This can be seen from the approximate formula for region III in Eq.~\eqref{eq:shortrangeapprox}, which shows that the fiducial cross section scales with $1/E_{\text{thres}}$. The rate per unit exposure in Eq.~\eqref{eq:masterform} is moreover proportional to $1/m_N = 1/(A\, m_n)$, such that there is no residual enhancement for large nuclei. This implies that the relative reach of the various experiments in this regime is determined by the ratio of their exposure over their threshold, assuming comparable backgrounds rates.

For $\mxbar=10$ GeV (Fig.~\ref{fig:nucleoncoupling_constituent10GeV}), the ADM nugget has a much smaller radius and thus does not suffer from form factor suppression for even intermediate masses. This means that the traditional, large exposure experiments perform best in most of the parameter space, except for very high $M_X$. Concretely, only for $M_X\gtrsim 10^{13}$ GeV a low threshold Si experiment with kg-year exposure would outperform the existing XENON1T limit, and even in this case a future liquid xenon experiment is still expected to set the best limit. In the intermediate regime for the constituent mass, 10 MeV $< \mxbar < 10$ GeV, the reach of both types of experiments is complementary: relatively low $M_X$ is covered by liquid xenon detectors, while the reach for high $M_X$ can be better for a low threshold experiment.\footnote{The reach shown for He and Si is reduced because we assume an upper bound on the detectable energy, 100 meV and 1 eV, respectively. This is an estimate for the final generation of detector concepts, with the lowest thresholds. As these campaigns will occur in stages, gradually lowering the energy range of interest, our projections are likely somewhat conservative for this benchmark. The upper bound on deposited energy also invalidates some of the analytic relations in Sec.~\ref{nuclearrecoilssection}; the necessary modifications are easily derived.}

\begin{figure}[t]
\begin{center}
\includegraphics[width=0.75\textwidth]{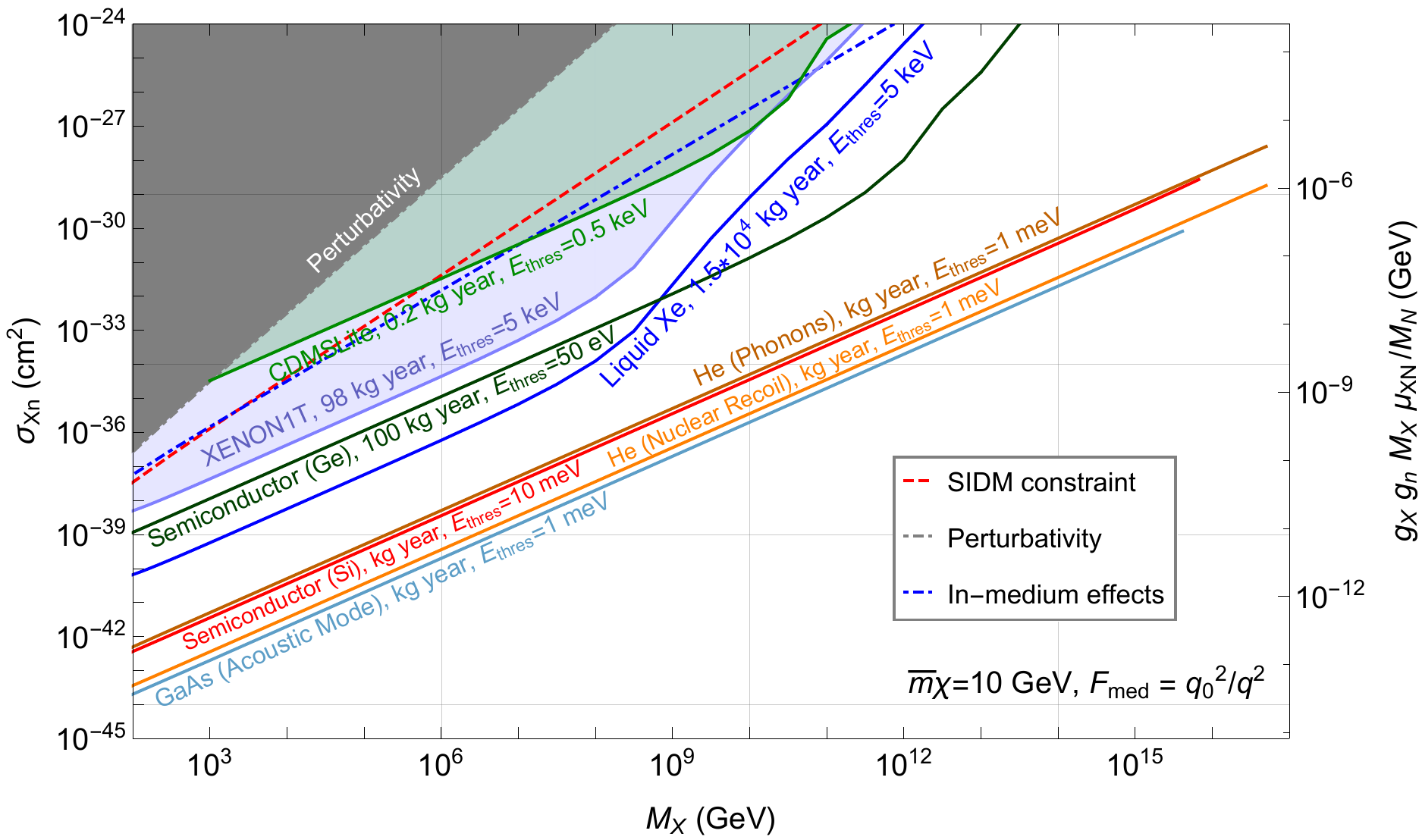}
\caption{Existing and projected reach of experiments described in Table~\ref{tab:nucleonexperiments} for light mediator, coupling to nucleons and nuggets with $\mxbar=$ 10 GeV constituents. At high $M_X$, the curves terminate when the expected number of ADM nuggets passing through the detector volume per year drops below one.  Above the dot-dashed blue line, the mediator receives important in-medium corrections inside the nugget. For the perturbativity, SIDM and in-medium correction bounds, we saturated the stellar cooling limits by fixing $g_n=10^{-12}$, assuming a mediator mass of 1 eV (See text for details.)}
\label{fig:nucleoncoupling_constituent10GeV_masslessmediator}
\end{center}
\end{figure}

The reach for a light mediator is shown in Fig.~\ref{fig:nucleoncoupling_constituent10GeV_masslessmediator}. ``Light'' here refers to a mediator whose mass is low compared to the typical momentum transfer in the scattering, meaning $m_\phi \ll m_N v_0$. The scattering is heavily biased towards the low momentum transfer regime, and a low threshold detector is always advantageous, regardless of the radius of the nugget (see Eq.~\eqref{eq:longrangeapprox}). In this scenario the phonon modes for helium and GaAs have the potential to outperform the high exposure experiments with several orders of magnitude in the full parameter space. For completeness, we also include a number of indirect constraints, for which we assumed the mediator mass $m_\phi$ to be 1 eV. For this mass value, the maximal coupling to the mediator to the SM nucleons is $g_n \sim 10^{-12}$, as derived from stellar cooling bounds \cite{Hardy:2016kme}. This constraint on $g_n$ is combined with several constraints on $g_\chi$ to demarcate regions that are either excluded via indirect constraints or must be handled with care.  For example, DM self-interactions via a light mediator, derived in Eq.~\eqref{eq:sidmlongrange}, exclude regions above the red dashed line, labeled as SIDM; this bound can be avoided if ADM nuggets are a subcomponent of the total DM density. There is also a region in Fig.~\ref{fig:nucleoncoupling_constituent10GeV_masslessmediator} which is a priori physical, but for which our approximations do not apply: above the dot-dashed blue line, corresponding to Eq.~\eqref{eq:screening}, the mediator mass receives important in-medium corrections which should be accounted for in the scattering rate. This effect is expected to weaken the constraints; for more details we refer to Sec.~\ref{sec:otherconstraints}.

\FloatBarrier

\section{Electron couplings\label{sec:electroncoupling}}

In this section, we consider scattering of ADM nuggets with electron couplings, including interactions mediated by dark photons. The leading existing electron-recoil limits are derived from XENON10 data \cite{Essig:2011nj,Essig:2012yx,Essig:2017kqs,Angle:2011th} and a recent surface run by the SENSEI collaboration \cite{Crisler:2018gci}; SENSEI had a factor of $\sim 10$ lower threshold, but substantially smaller exposure. A comparable result exists from the SuperCDMS collaboration \cite{Agnese:2018col}.  Near future (1-5 years) detectors will have enhanced sensitivity to electron recoils, primarily by pushing for higher exposure, and in some cases lower thresholds. This includes semiconductor targets (SENSEI \cite{Battaglieri:2017aum,Crisler:2018gci}, DAMIC-K \cite{Settimo:2018qcm}, SuperCDMS \cite{Battaglieri:2017aum}), scintillators \cite{Derenzo:2016fse}, graphene (PTOLEMY \cite{Hochberg:2016ntt,Baracchini:2018wwj}) and xenon (LBECA \cite{Battaglieri:2017aum}).  As an example we take a silicon semiconductor with a kg-year exposure with a two electron threshold, as a future projection for SuperCDMS, SENSEI or DAMIC \cite{Battaglieri:2017aum}. Finally, there are slightly longer term prospects for even lower threshold detectors, such as superconducting aluminum \cite{Hochberg:2015fth,Hochberg:2015pha}, Dirac materials \cite{Hochberg:2017wce}, and polar materials (e.g.~GaAs, sapphire) \cite{Knapen:2017ekk,Griffin:2018bjn}. As representative examples we consider a superconducting target and, in the case of a dark photon mediator, a polar material (GaAs). All (proposed) experiments we consider are listed in  Table~\ref{tab:electronexperiments}, along with their (assumed) thresholds and exposure. Just as we did for the nucleon couplings, we have assumed aspirational exposures for the more long term proposals; rescaling the projected limits for different assumptions is trivial.

 \newcolumntype{C}[1]{>{\centering\let\newline\\\arraybackslash\hspace{0pt}}m{#1}}
\begin{table}[h]

\begin{tabular}{p{4.8cm}C{4cm}C{2.5cm}C{2.5cm}C{2cm}}\hline
Experiment&Exposure (kg-year)&Threshold&Timeline&$N_{\mathrm{events}}$\\\hline
XENON10 &0.041&$15$ eV&existing&BD\\
SENSEI &$5.2\times 10^{-8}$&$8.3$ eV&existing&\rm{BD}\\\hline
Si semiconductor&1&4.7 eV&in progress&3\\
superconductor &1&$1$ meV&R\&D needed&3\\
polar material ($A'$ mediator)&1&$30$ meV&R\&D needed&3\\
\hline
\end{tabular}
\caption{Overview of assumed exposure and approximate thresholds for the existing and future experiments considered in this work, for electron-nugget couplings. $N_{\mathrm{events}}$ refers to the number of expected signal events that were assumed to estimate the 90\% exclusion limits. (See text for details.)  ``BD'' refers to a bin-dependent analysis, as described in the text. \label{tab:electronexperiments}}
\end{table}

\subsection{Electron recoils}

For reference, we summarize the dynamic structure factors for electron targets appearing in Eq.~\eqref{eq:masterform}.  The dynamic structure factor, $S(\textbf{q},\omega)$, in semiconductors and liquid xenon depends strongly on the electron potential in the medium.  The electrons in a superconductor can be approximated by a non-interacting Fermi liquid at zero temperature, and the impact of Pauli blocking must be accounted for.

\subsubsection{Ionization in Atomic Targets}
In this process, the DM ionizes an electron in one of the outer shells of the target atom, Xe in the case at hand. The residual kinetic energy of the electron is deposited in the form of secondary electrons or (unobserved) scintillation photons. The corresponding structure function is
\begin{equation}
S(q,\omega)=\frac{1}{4}\sum_{n,\ell} \frac{1}{\omega-|\epsilon_{n,\ell}|}|f_{n,\ell}(k',q)|^2
\label{eq:IonSF}
\end{equation}
with $\epsilon_{n,\ell}$ the binding energy of the atomic level $(n,\ell)$, $k'\equiv \sqrt{2 m_e (\omega-|\epsilon_{n,\ell}|)}$ and the ionization form factor. The form factor $f_{n,\ell}(k',q)$ is non-trivial and must be computed from the atomic wave functions and the wave function of the outgoing, unbound electron, as detailed in Refs.~\cite{Essig:2011nj,Essig:2012yx,Lee:2015qva,Essig:2017kqs}. We compute the structure function and the electron yield as a function of the recoil energy with \verb+QEdark+ code \cite{Essig:2012yx,Essig:2017kqs}, which supplies $f_{n,\ell}(k',q)$ for Xe. Since the rate is dominated by the $4d^{10}$ and $5p^{6}$ shells, we neglect the remaining orbitals in our estimates. We further bin the differential rate according to the expected number electrons and require the signal to be below the XENON10 data \cite{Angle:2011th} at 90\% confidence level in each bin. The above approximations reproduce the limits for light, elementary dark matter in \cite{Essig:2017kqs} to within $\sim 30\%$.

\subsubsection{Electronic Transitions in Semiconductors}

In semiconducting targets, such as silicon and germanium, the gap between a valence and conduction band is 1.1 eV and 0.67 eV respectively, which corresponds to the lower bound on the energy threshold. In practice, the threshold is often taken to be the energy needed to produce two or three electrons, in order to suppress the dark count rate and/or other low energy backgrounds. The structure function is 
\begin{equation}
S(q,\omega)=\frac{1}{n_T}\sum_{i,i'}\sum_{G'}\int_{BZ}\! \frac{d^3k}{(2\pi)^3} |f_{i,k\to i',k',G'}|^2\delta\left(E_{ i',k'}-E_{i,k} -\omega\right)
\end{equation}
where $n_T$ is the electron number density and the $i,i'$ label the initial and final electronic bands, with energy $E_{i,k}$ at momentum $k$. The sum over $G'$ is over the reciprocal lattice and the integrals over the initial and final electron momenta $k,k'$ are over the first Brillouin zone. The transition matrix element $f_{i,k\to i',k',G'}$ is determined from the electron wave functions and has been evaluated numerically \cite{Essig:2015cda} with the \verb+Quantum EXPRESSO+ \cite{0953-8984-21-39-395502} package, and subsequently tabulated as part of the  \verb+QEdark+ code \cite{Essig:2015cda,Essig:2012yx}, which we use for our calculations. (See \cite{Lee:2015qva} for a semi-analytic approach.)
 
 To convert the energy deposited by the DM to the number of ionization electrons ($n_e$), we assume the same linear relationship as in \cite{Essig:2015cda}. For the existing bounds, we recast the latest SENSEI limit \cite{Crisler:2018gci}. Since this was a surface run, the one and two electron bins have a large amount of background events and we set a limit only with the $n_e\gtrsim 3$ and $n_e\gtrsim 4$ electrons selections, which respectively correspond to a threshold of 8.3 eV and 11.9 eV. In these two signal regions, 132 and 1 event(s), respectively, were observed and the bounds at 90\% confidence level are computed by conservatively assuming that all observed events were signal. For each point in the parameter space, we take the strongest of these two limits.  For the future reach of a larger exposure, underground detector, a negligible background is assumed for $n_e\gtrsim2$, which corresponds to an effective threshold of 4.7 eV; as a benchmark, we assume an exposure of 1 kg-year.

\subsubsection{Scattering in Superconductors}
\label{sec:supercond}
In metal targets, DM can scatter off of the quasi-free valence electrons \cite{Hochberg:2015pha,Hochberg:2015fth}, which have a typical Fermi velocity, $v_F \sim 10^{-2}$. In the superconducting phase, a small gap on the order of an meV develops above the Fermi surface, which forbids scattering processes with energy depositions less than the width of the gap.  For energy depositions sufficiently above a meV, the existence of the gap can be neglected and the scattering becomes identical to that in a free Fermi gas, at zero temperature.  In this case, the dynamic structure factor is  
\beq
\begin{aligned}
S(q,\omega) &= \frac{m_e v (2 m_e \mu - \xi^2)}{2 \pi q\ n_T} \Theta(\sqrt{2 m_e \mu} - \xi) \\
\xi& = \textrm{Max}\left[\sqrt{2 m_e (\mu - \omega)}, \frac{m_e}{q} \left(\omega - \frac{q^2}{2 m_e} \right) \right]
\end{aligned}
\eeq
where $m_e$ is the mass of an electron, $\mu$ the chemical potential and $n_T$ is the electron number density. Note that for superconducting aluminum, $\mu\approx11.7$ eV. The structure function is derived in greater detail in Appendix \ref{zeroTsuperconductor}. For the future reach, we assume a  meV threshold, kg-year exposure and negligible irreducible backgrounds. 

If the particle mediating the scattering is a kinetically mixed dark photon, the scattering is subject to screening by the valence electrons of the metal. In practice the rate is suppressed by the Thomas-Fermi screening length, which implies a reduction of the reach of several orders magnitude, depending on the momentum transfer. For our rate calculation we include the full momentum dependent correction to the dark photon propagator, following the discussion in \cite{Hochberg:2015fth}.  

\subsection{Polar materials}

The detector with a polar material target proposed in \cite{Knapen:2017ekk} can be described as an ultra-low threshold calorimeter for single, athermal phonon excitations. Since polar materials are insulators or semiconductors, the 1-10 meV threshold needed to detect single phonons is well below the band gap of the material. Unlike the examples discussed above, the detector is therefore not optimized for detecting electron excitations directly, but it is nevertheless still possible to excite phonons through the coupling to the inner shell electrons \cite{Griffin:2018bjn}. For a mediator coupling to electrons, a superconducting target tends to perform somewhat better than a polar material target. The kinetically mixed dark photon mediator is however an important exception due to the screening effect described in the previous section, which strongly limits the reach of superconductors for this scenario. For an insulator (e.g.~sapphire) or a semiconductor (e.g.~GaAs) this issue does not arise. 

A defining feature of polar materials is that the optical phonon modes of the crystal correspond to the coherent oscillation of an electric dipole in each unit cell, which implies an enhanced coupling to a kinetically mixed dark photon. In the limit where the dark photon is massless, the ADM nugget can be thought of as carrying a small electric charge. In this case the dynamic structure factor is \cite{Knapen:2017ekk,Griffin:2018bjn}
\begin{equation}\label{eq:frolichstructure}
S(q,\omega)=  \frac{2 q^2}{e^2} \frac{\omega_{LO}}{n_T}\left(\frac{1}{\epsilon_\infty}-\frac{1}{\epsilon_0}\right)\delta(\omega_{LO} - \omega),
\end{equation} 
taking $\bar \sigma_0$ in Eq.~\eqref{eq:modeleq} (for the massless case), $g_e=e$ and $g_t = \kappa e$ with $e$ the electron charge and $\kappa$ the kinetic mixing parameter. $\epsilon_0$ and $\epsilon_\infty$ are the low and high frequency dielectric constants respectively, and $\omega_{\text{LO}}$ is the energy of the longitudinal optical phonon in the zero momentum limit. The reference momentum is taken to be $q_0 = \alpha m_e$.\footnote{Note that our definition of the reference cross section $\bar \sigma_0$ differs with a factor of 4 from the definition in \cite{Knapen:2017ekk,Griffin:2018bjn}.} We note that the above equation is an analytic approximation, valid only for isotropic crystals like GaAs. The general structure function is more complicated and must be evaluated numerically; see \cite{Griffin:2018bjn} for details. Since the numerical treatment is rather computationally intensive, we restrict ourselves to the analytic formula in Eq.~\eqref{eq:frolichstructure} for GaAs in this work. As in Sec.~\ref{sec:coherentmodes}, we require $q< 1$ keV, to ensure the validity of the phonon description.

\subsection{Results}


We now turn to a quantitative comparison of the numerical results. As in the case of coupling to nucleons, the key quantity is the radius of ADM nugget, given by
\beq
R_X = \ \left(\frac{9\pi}{4}\frac{M_X}{\bar{m}_\chi^4}\right)^{1/3}.
\label{eq:radiusformula2}
\eeq
In our reach estimate, we again fix $\mxbar$ to two benchmark values and vary the ADM nugget mass $M_X$. As for DM coupling to nuclei, the scaling of a detector's sensitivity falls into three unique regions (see Fig.~\ref{fig:hetwokinks}), depending on how the momentum transfer compares to the inverse radius. In Region I, the maximum allowable momentum transfer, $q_\text{max}$, is smaller than the inverse radius. The scattering of a composite nugget is therefore fully coherent and also indistinguishable from that of an elementary DM particle. In Region II, while $q_\text{max} R_X \gtrsim 1$, the lowest momentum transfer to which the detector is sensitive, $q_\text{min}$, is still smaller than $1/R_X$. The scattering is therefore still coherent, but only in part of the phase space. In Region III, $q_\text{min} R_X \gtrsim 1$ and the scattering is never coherent and thus the rate is strongly suppressed.

 The transitions between Region I and II and Region II and III occur roughly when the inverse radius is small compared to the maximal momentum transfer and the experimental momentum transfer threshold, i.e.
\beq\label{eq:electrontransionregions}
M_X \sim \begin{cases}  3 \times 10^6 \GeV\times \left(\frac{\mxbar}{10 \MeV}\right)^4\times\left(\frac{8 \keV}{q_\text{max}}\right)^3 &\quad \text{transition between Region I and II} \\ \\
10^{9} \GeV \times\left(\frac{\mxbar}{10 \MeV}\right)^4\times\left(\frac{\eV}{E_\text{thres}}\right)^3 &\quad \text{transition between Region II and III} \end{cases},
\eeq
where $v_0$ is the typical DM velocity in the Milky Way and $q_\text{max} \approx 2\alpha m_e$, as explained below; this is significantly smaller than in the nuclear recoil case. Notice that, at least for atomic ionization and semiconductor detectors, Region II is fairly narrow, as $q_\text{max}$ is numerically close to $q_{\text{min}}\approx E_\text{thres}/v_0$.

The structure functions are substantially more complicated than those for nucleon couplings in Sec.~\ref{sec:nucleoncoupling}, but we can still extract the rough scaling of the scattering rate by making a number of additional approximations: if the electron were free, the structure function would simply be Eq.~\eqref{eq:nucleonxsec}, except with the mass of the electron instead of mass of the nucleus and $A=1$. The effect of the electron being in a bound state is two-fold: first,  the energy transferred to the electron, $E_e$, is now related to the momentum transfer via
\beq
E_e =  \bfq \cdot \bfv - \frac{q^2}{2\mu_{X N}} \approx \bfq \cdot \bfv \, ,
\eeq
found by imposing energy conservation on the DM-electron-nucleus system. This implies the relation
\begin{equation}\label{eq:electronqmin}
q_{\text{min}}=E_{\text{thres}}/v_0
\end{equation}
between the minimum possible momentum transfer ($q_{\text{min}}$) and the detector threshold ($E_{\text{thres}}$).
Secondly, the electron is not only not at rest, but its velocity in the atom, $v_e$, is larger than the mean DM velocity. This implies that $q_\text{max} \approx 2 m_e v_e$, where $v_e \approx \alpha\approx 10^{-2}$, and we expect that the cross sections for scattering via a heavy mediator roughly scales as
\beq \label{eq:shortrangeapproxElectron}
\sigma^{\text{heavy}}_{ER}&\sim& \overline \sigma_0 \times \begin{cases}\frac{q^2_{\text{max}}}{4v_0^2 m_e^2}= \frac{v^2_e}{v_0^2}& \text{Region I} \\
\frac{1}{R_X^2}\frac{2}{ v_0^2 m_e^2}& \text{Region II}\\
\frac{1}{R_X^4}\frac{ 2}{q^2_{\text{min}} v_0^2 m_e^2}=\frac{1}{R_X^4}\frac{2}{m_e^2 E^2_{\text{thres}}  }& \text{Region III}\end{cases}\,\,,
\eeq
while for a light mediator, we expect the cross section to scale as
\beq \label{eq:longrangeapproxElectron}
\sigma^{\text{light}}_{ER}&\sim& \overline \sigma_0 \times q_0^4\times \begin{cases}
\frac{1}{4q_\text{min}^2 v_0^2 m_e^2}=\frac{v^2_e}{v_0^2}\frac{1}{m_e^2E^2_\text{thres}  }& \text{Region I \& II} \\ \\
\frac{1}{R_X^4}\frac{1}{v_0^2 m_e^2 q_{\text{min}}^6}=\frac{1}{R_X^4}\frac{v_e^6}{v_0^2}\frac{1}{m_e^2 E_\text{thres}^6}& \text{Region III}\end{cases}\,\,\,\,.
\eeq
Note that while these electron recoil cross sections have the same scaling dependence on $q_\text{max},\, q_\text{min}$ and $R_X$ as for the nuclear recoil case, the dependence on $E_\text{thres}$ is altered, following Eq.~\eqref{eq:electronqmin}. These relatively crude approximations broadly reproduce the scaling behavior of the reach of the various detector types, but their quantitative accuracy is not always guaranteed, as more subtle but numerically important in-medium effects (e.g.~Pauli blocking) are not accounted for. However, these in-medium effects can also sometimes be included in a similarly rough manner; when appropriate, we will provide an estimate of the consequences of these effects on specific detector limits in our discussions below.  In our result plots, we always present the full, numerical results, as outlined in the preceding sections.

\begin{figure}[p]
\begin{center}
\includegraphics[width=0.75\textwidth]{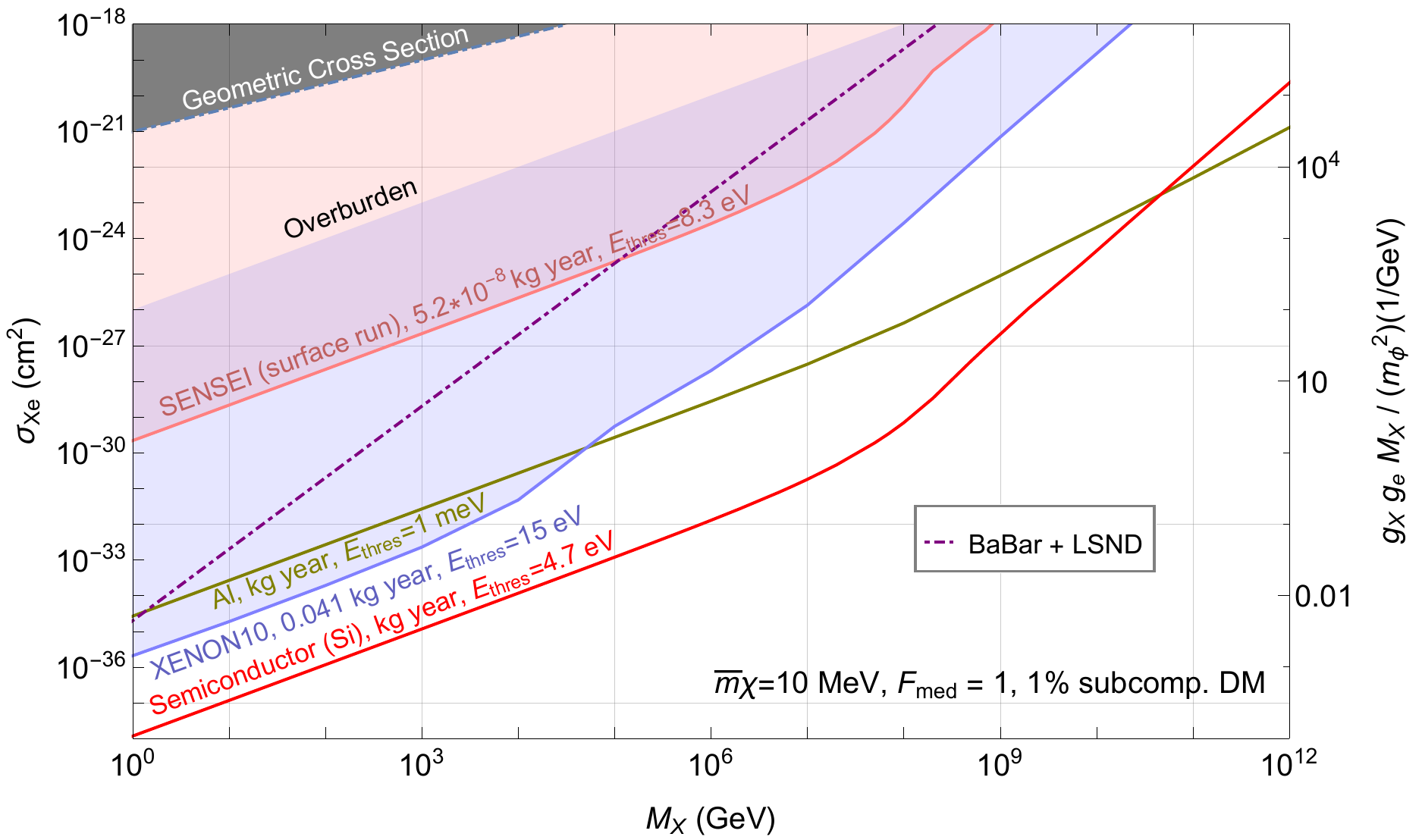}
\caption{
Existing and projected reach of experiments in Table~\ref{tab:electronexperiments} for heavy mediator interacting with  electrons and nuggets with 10 MeV constituents.  We assume nuggets make up 1\% of the total DM density to evade SIDM constraints; the dot-dashed purple line is an intensity frontier constraint on a 10 MeV mediator; see Sec.~\ref{sec:MedInteractionSM}. The grey area marks where geometric cross section is exceeded.}
\label{fig:electroncoupling_constituent1MeV}
\end{center}
\end{figure}

\begin{figure}[p]
\begin{center}
\includegraphics[width=0.75\textwidth]{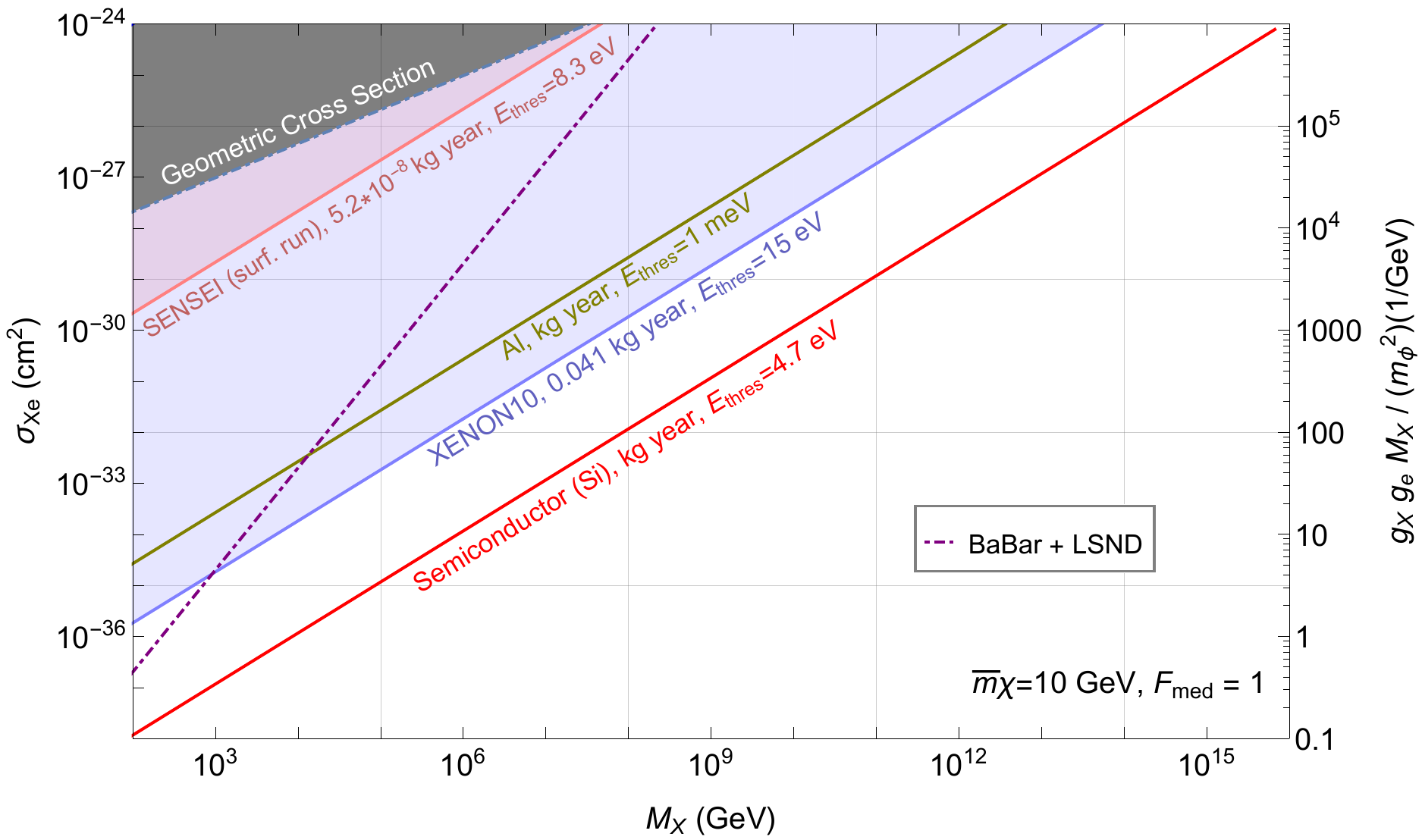}
\caption{Existing and projected reach of experiments in Table~\ref{tab:electronexperiments} for heavy mediator coupled to electrons and nuggets with 10 GeV constituents. The dot-dashed purple curve is an intensity frontier constraint on a 10 MeV mediator, see Sec.~\ref{sec:MedInteractionSM}. The grey area marks where geometric cross section is exceeded.
}
\label{fig:electroncoupling_constituent10GeV}
\end{center}
\end{figure}

The reach for a heavy mediator is shown in Figs.~\ref{fig:electroncoupling_constituent1MeV} and \ref{fig:electroncoupling_constituent10GeV} for $\mxbar=$ 10 MeV and $\mxbar=$ 10 GeV respectively. The grayed region again indicates cross sections larger than geometric, which are unphysical. For electron couplings there is a mild overburden effect for underground experiments, as argued in Appendix \ref{app:overburden}. The parameter space where this is relevant is however fully covered by the SENSEI surface run. The dot-dashed purple line refers to the constraint in Eq.~\eqref{eq:lepbound}, which represents the intensity frontier bounds on the mediator particle, specifically from BaBar and LSND (see Sec.~\ref{sec:MedInteractionSM} for details).

When in Region I, i.e. when the ADM nuggets are quite small, low threshold detectors (superconductors in the present context)  do not have an advantage over their counterparts with higher thresholds: comparing a 1 eV threshold semiconducting experiment to a meV threshold superconducting experiment only shows the effect of Pauli blocking, which weakens the constraints by approximately three orders of magnitude.  The initially surprising appearance of Pauli blocking can be understood in the following way: While one might naively expect that the rate for DM scattering in superconductors should be dominated by electrons deep in the Fermi sea (because the nugget carries plenty of kinetic energy), the rate is instead dominated by energy deposits near the detector threshold.  The physical reason is that scattering of heavy, slow-moving dark matter off of a fast electron typically imparts a momentum transfer twice the initial electron momentum. For the electrons deep in Fermi sea, this does not suffice to knock the electron across the Fermi surface, and as such the process is Pauli blocked. Thus one can only scatter off electrons near the Fermi surface, which substantially reduces the number of available scattering centers for the ADM nugget. 

Of the existing constraints for $\mxbar=10$ MeV (Fig.~\ref{fig:electroncoupling_constituent1MeV}), the XENON10 limit is substantially stronger than the SENSEI limit, due to its much larger exposure. The 1, 2 and 3 electron bins for XENON10 however have a substantial amount of background, such that the reach at low $M_X$ is primarily driven by bins with 4 or more electrons, effectively raising the threshold to $\sim 60$ eV. With Eq.~\eqref{eq:electrontransionregions}, this explains why the limit from XENON10 starts to deteriorate at lower $M_X$, as compared to the SENSEI limit. Focusing next on the reach of future semi- and superconductor experiments, note that both have the same  $q_\text{max}$, as it is set by twice the typical electron momentum. Therefore, both detectors transition between Region I and Region II  around $M_X\gtrsim3\times 10^6$ GeV, and the difference in their reach at low nugget masses can be understood completely through the existence of Pauli blocking in superconductors.  Moreover, the semiconductor detector rather quickly transitions from Region II to Region III, due the numerical coincidence that $q_\text{min}$ is less than an order of magnitude smaller than $q_\text{max}$ for this detector. This transition is delayed for the low threshold, superconducting device, such that it can eventually compensate for the suppression due to Pauli blocking. This is what occurs in Fig.~\ref{fig:electroncoupling_constituent1MeV}, where superconductors begin to dominate over semiconductors around $M_X \approx 10^{11}$ GeV.  

For more compact nuggets, i.e. with $\mxbar=$ 10 GeV, the ADM nuggets behave like point particles for all detectors, and there is no form factor suppression. The semiconductor (or ionization) experiments therefore always dominate, since they do not suffer from Pauli blocking. The reach is therefore primarily exposure-driven, as is shown in Fig.~\ref{fig:electroncoupling_constituent10GeV}. A future noble liquid detector with sensitivity to ionization signals and more exposure than XENON10 (e.g.~LBECA) could therefore be an interesting alternative to search for ADM nuggets, or heavy DM with electron couplings more generally. That is, provided that the backgrounds can be kept low, a caveat which applies to all future proposals we consider. 

\begin{figure} [p]
\begin{center}
\includegraphics[width=0.75\textwidth]{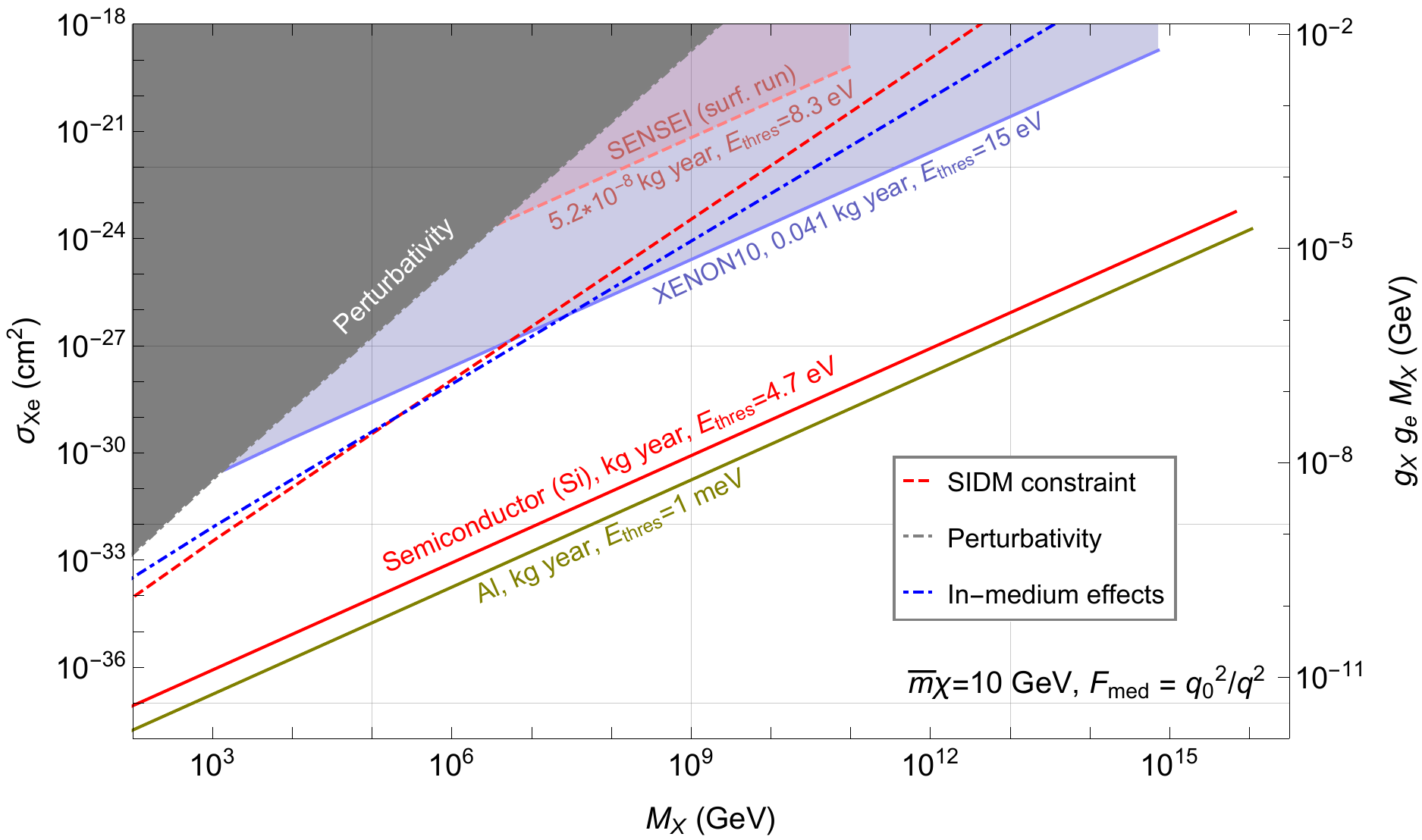}
\caption{Reach curves of experiments in Table~\ref{tab:electronexperiments} for light scalar mediator. The perturbativity, in-medium effects and SIDM constraints are drawn assuming $g_e = 10^{-15}$, consistent with stellar cooling bounds for $m_\phi=1$ eV. Pauli blocking is present only in superconductors.
}
\label{fig:electroncoupling_constituent10GeV_masslessmediator}
\end{center}
\end{figure}

\begin{figure} [p]
\begin{center}
\includegraphics[width=0.75\textwidth]{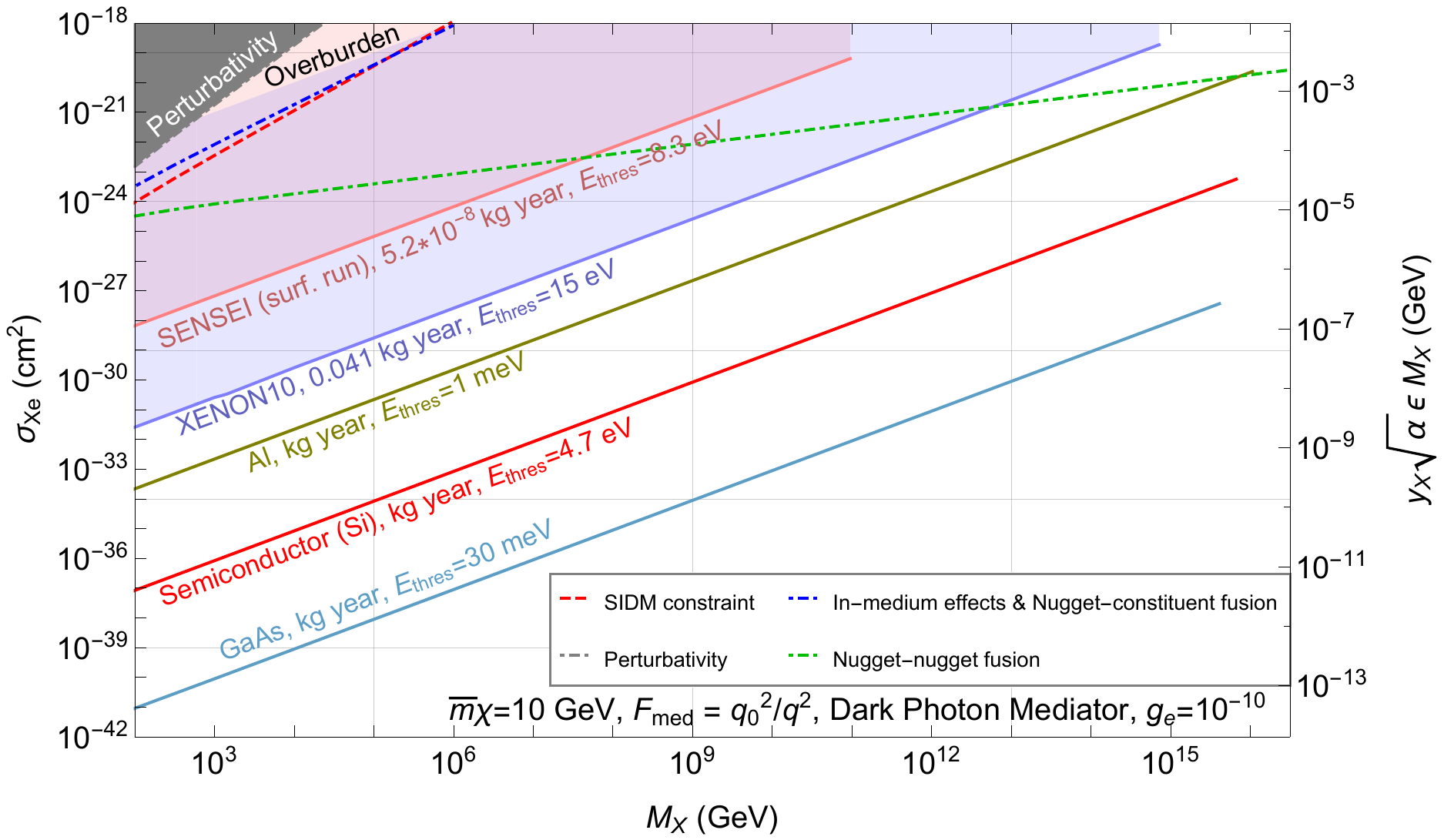}
\caption{Reach curves for experiments listed in Table~\ref{tab:electronexperiments} for dark photon mediators. All non-direct detection constraints, denoted by dot-dashed lines, are drawn assuming $g_e = \kappa e = 10^{-10}$, consistent with stellar cooling bounds for $m_{A'}=10^{-2}$ eV. Early universe nugget formation constraints how large $g_\chi$ can be, marked with blue and green dot dashed curves.}
\label{fig:electroncoupling_constituent10GeV_darkphotonmediator}
\end{center}
\end{figure}

The reach for a light scalar mediator and for a light, kinetically mixed dark photon is shown in Figs.~\ref{fig:electroncoupling_constituent10GeV_masslessmediator} and~\ref{fig:electroncoupling_constituent10GeV_darkphotonmediator} respectively; both are for point-like nuggets, with $\mxbar = 10\GeV$.  As in the nuclear recoil case, a mediator is considered  ``light'' if its mass is small compared to the typical momentum transfer, in this case $m_\phi \ll \alpha m_e$.  For the scalar mediator case, we take the mediator mass to be $m_\phi=$ 1 eV, which corresponds to a maximal coupling to electrons of $g_e \sim 10^{-15}$ from stellar cooling constraints \cite{Hardy:2016kme}.  For the vector mediator case, we fix $g_e = 10^{-10}$, which is consistent with stellar cooling bounds for a dark photon with mass $m_{A'}=10^{-2} \mbox{ eV}$ \cite{An:2013yfc}.  This constraint is combined with the constraints from perturbativity ($g_\chi \lesssim 1$) and dark matter self-interactions in Eq.~\eqref{eq:sidmlongrange}. We also include the lines that demarcate regions that are a priori physical, but for which our approximations do not hold. In particular, in the region above the dot-dashed blue line, defined by Eq.~\eqref{eq:screening}, the mediator mass receives important in-medium corrections, as the charge density of the fermions can dramatically alter the light mediator's potential. Lastly, note that the maximum nugget mass that GaAs, silicon semiconductor and aluminum superconductor detectors can probe with a kg-year exposure are slightly different due to the difference in the detector volumes.

In direct detection experiments,  the scattering rate for interactions via a light mediator is always dominated by low momentum transfer, due to the $q_0^2/q^2$ form factor in Eq.~\eqref{eq:medformfactor} and so low threshold experiments always perform best, as long as Pauli blocking or in-medium screening effects are not too severe. Focusing first on scalar mediators (Fig.~\ref{fig:electroncoupling_constituent10GeV_masslessmediator}), the existing constraints are provided by XENON10 and the surface run of  SENSEI. For the future reach, superconductors dominate over semiconductors, despite the Pauli blocking suppression, as the rate per unit exposure is proportional to $1/q_\text{min}^2 \sim 1/E_\text{thres}^2$.  

There are several important differences when considering a vector mediator (Fig.~\ref{fig:electroncoupling_constituent10GeV_darkphotonmediator}). For polar crystals (e.g.~GaAs),  the dark photon has a strong coupling to the optical phonons in material, which enhances their sensitivity. Superconductors on the other hand pay a large penalty in reach due to the screening by the valence electrons, as discussed in Sec.~\ref{sec:supercond}. For indirect constraints, we also demarcate, with a dot-dashed green line, where the dark Coulomb barrier would forbid nuggets from successfully undergoing fusion in the early Universe, as given in Eq.~\eqref{eq:gchifusions}. Depending on whether the last fusion reaction to freeze out is between two large nuggets or a large nugget and a free constituents or small nugget affects the mass of the maximum nugget that can be synthesized \cite{Hardy:2014mqa, Gresham:2017cvl} and so we choose to plot both of these constraints.

\section{Conclusions\label{sec:conclusions}}

We have considered the detection of nuggets of Asymmetric Dark Matter, comparing and contrasting low threshold but low exposure proposed experiments (such as superconductors, superfluid helium, polar materials and semi-conductors) with higher threshold but larger exposure experiments (notably the traditional noble liquid xenon experiments).  We found that both types of experiments have a role to play:
\begin{itemize}
\item Larger experiments are more successful at accessing compact and lighter nuggets that interact with the Standard Model via a heavy mediator;
\item Smaller, but lower threshold, experiments can dominate for less dense nuggets and for nuggets which interact with the Standard Model via a longer range mediator.
\end{itemize}

It has been shown that large bound states of Asymmetric Dark Matter arise quite generically in the presence of a sufficiently strong attractive force, and their astrophysical and cosmological evolution can be quite distinct from other known DM candidates. It is therefore important to pursue complimentary search strategies for constraining these intriguing dark matter candidates.

There are number of possible future directions: Firstly, in our analysis we have not considered the likely possibility that the ADM nuggets follow a non-trivial mass distribution, which can be bimodal in the case of a bottleneck in the formation history \cite{Gresham:2017cvl}. In such a scenario, there could be a signal from larger nuggets in one or more low threshold experiments, simultaneously with a signal from the smaller nuggets in the traditional large noble liquid detectors. Moreover it could be interesting to revisit some of the prior studies of the differential energy spectrum \cite{Hardy:2015boa,Butcher:2016hic} by including low threshold detectors, as well as correlating the spectrum with the formation history of the ADM nuggets.

\FloatBarrier

\section*{Acknowledgments}
We thank Marat Freytsis, Keisuke Harigaya, Tom Melia, Matt Pyle, Surjeet Rajendran, Harikrishnan Ramani, Diego Redigolo, Tomer Volansky and Tien-Tien Yu for useful discussions, and Tomer Volansky and Tien-Tien Yu for assistance with the QEdark package.  KZ is supported by the DoE under contract No. DE-AC02- 05CH11231.  AC and KZ are supported by the Quantum  Information Science Enabled Discovery (QuantISED) for High Energy Physics (KA2401032).  DMG is funded under NSF Grant 32539-13067-44-PHHXM and DOE Grant 041386-002. Part of this work was performed at the Aspen Center for Physics, which is supported by National Science Foundation grant PHY-1607611. DMG thanks the Aspen Center for Physics and Kavli Institute for the Physics and Mathematica of the Universe (IPMU) for the hospitality shown while this work was being completed. The work by SK was supported in part by the LDRD program of LBNL under contract DE-AC02-05CH11231, and by the National Science Foundation (NSF) under grants No.~PHY-1002399 and PHY-1316783. SK also acknowledges support from DOE grant DE-SC0009988 and from the Kavli Institute for Theoretical Physics, supported in part by the National Science Foundation under Grant No.~NSF PHY-1748958, where part of this work was performed. 

\appendix
\section{Material overburden}
\label{app:overburden}
In this appendix we calculate the material overburden due to the rock above the experiments. The relevant quantity is the energy lost per unit of distance that was traveled through the material in question. For an ADM nugget coupling to hadronic matter and with velocity $v$, the average energy loss per collision is
\begin{align}\label{eq:dEdx}
\left\langle\Delta E\right\rangle_{v_0}&=\frac{1}{\overline\sigma_{XN}}\int\! dE\, dq\; E\,|F_X(q)|^2 |F_{\text{med}}(q)|^2\frac{d\bar\sigma_{XN}}{d E\, dq}\\
&=\frac{1}{\overline\sigma_{XN}}\frac{A^2}{4 m_N m_n^2 v_0^2}\int\!  dq\; q^3\,|F_X(q)|^2 |F_{\text{med}}(q)|^2
\end{align}
with
\begin{align}
\overline\sigma_{XN}&=\int\! dE\, dq\; |F_X(q)|^2 |F_{\text{med}}(q)|^2\frac{d\bar\sigma_{XN}}{d E\, dq}\\
&=\frac{A^2}{2 m_n^2 v_0^2}\int\!  dq\; q\,|F_X(q)|^2 |F_{\text{med}}(q)|^2,
\end{align}
where we took the $M_X\gg m_N$ limit. The average energy loss per unit of length is thus
\begin{equation}\label{eq:overburden}
\left\langle \frac{d E}{dx}\right\rangle_{v_0} \approx \text{Min}\big[\,n_T \overline\sigma_{XN},n_T^{1/3}\,\big]\times \left\langle \Delta E\right\rangle_{v_0},
\end{equation}
with $n_T$ the number density of the material. The second term in Min accounts for the possibility that the nugget scatters off every nucleus it meets, which occurs if the cross section is sufficiently large \cite{Bramante:2018qbc}. The presence of the form factor implies that the overburden depends on the nugget radius. For large nuggets, the probability of transferring a sizable amount of momentum is suppressed, these nuggets are much less likely to get stopped, as compared to a point-like DM with a comparable mass and cross section.

For ADM nuggets coupling to electrons through a light mediator, we estimate the energy loss with the Lindhard-Scharff formula for electronic energy loss \cite{Lindhard:1961zz}. For a singly ionized atom with speed $v$ and atomic number $Z_1$, traveling the through medium with atomic number $Z_2$, the average energy loss per unit distance is
\begin{align}
\left\langle\frac{dE}{dx}\right\rangle&\approx 8\pi n_2 e^2 a_0\frac{Z_1 Z_2}{(Z_1^{2/3}+Z_2^{2/3})^{3/2}} \frac{v}{v_0}.
\end{align}
Here $n_2$ the number density of the target, $e$ the electron charge, $a_0$ the Bohr radius and $v_0\approx \alpha$ the typical velocity of a bound electron.
This prescription has shown to be in reasonably good agreement with the data for a variety of projectile and target ions \cite{LAND1978235}. At low velocity the elastic atomic recoil can be of comparable importance \cite{Lindhard:1961zz}, but since we only attempt an order of magnitude estimate here, we neglect this contribution. To estimate the overburden effect, we take the earth's crust to consist primarily of silicon $Z_2=14$ with a mean density of $n_2=2.7 \,\mathrm{g/cm^3}$. We take $Z_1=1$, since the nugget does not have an electron cloud as long as its total effective electric charge is $\lesssim 1$. Because of the stringent stellar cooling bounds on $g_e$, this is always satisfied in our parameter space. For the light mediator case, we then estimate the energy loss per unit distance as
\begin{align}
\left\langle\frac{dE}{dx}\right\rangle&\approx 8\pi n_2 e^2 a_0 N_X g_\chi g_e \frac{v}{v_0}\approx 120 \times \frac{\mathrm{MeV}}{\mathrm{cm}}\times (N_X g_\chi g_e)^2 \times \frac{v}{10^{-3}}.
\end{align}
where we identified the dimensionless parameter $N_X g_\chi g_e$ with the electric charge in the Lindhard formula. (The energy loss in a dielectric is proportional to the square of the electric charge, see e.g.~\cite{Lindhardreview}.) Requiring that the total energy loss remains smaller than the kinetic energy implies
\begin{equation}
\sigma_{Xe}\ll 10^{-24} \,\mathrm{cm}^2 \times \left(\frac{M_X}{\mathrm{GeV}}\right),
\end{equation}
which is always satisfied for a scalar mediator in the parameter space we consider. For a vector mediator, there is a sliver of parameter space where this is not satisfied; however, this region is constrained by the SENSEI surface run. Finally, for a heavy mediator, the typical energy loss per collision is estimated to be $\langle \Delta E\rangle_0\approx \frac{1}{2}\alpha^2 m_e \approx 10$ eV. Using Eq.~\eqref{eq:overburden}, the ADM nugget would only get stopped in the earth's crust if 
\begin{equation}
\sigma_{Xe}>  10^{-26} \,\mathrm{cm}^2 \times \left(\frac{M_X}{\mathrm{GeV}}\right) \qquad \mathrm{and}\qquad M_X\lesssim 10^{11} \,\mathrm{GeV}
\end{equation}
which is always satisfied for nuggets with 10 GeV constituents, but not for nuggets with 10 MeV constituents; however, the region where overburden is important is always constrained by the SENSEI surface run. Note that for both the heavy and light mediators with electron couplings we have implicitly treated the ADM nuggets as point particles, which is a good approximation for low mass ADM nuggets, which is where the overburden effect is relevant.

\section{Zero temperature limit for superconductors}
\label{zeroTsuperconductor}
From \cite{Reddy:1997yr}, $S(E_D,q)$ can be calculated by
\beq
S(E_D,q) &= \frac{m_e v}{\pi q n_T} \int^\infty_{p_-} dp_2 \ p_2 f(E_2) (1 - f(E_4)),
\eeq
where $n_T$ is the number density of target, $p_2$ is the momentum of incoming electron, $p_4$ is the momentum of the outgoing electron and $f(E)$ is the Fermi-Dirac distribution:
\beq
\begin{aligned}
p_- = \frac{m_e}{q}\left(E_D - \frac{q^2}{2 m_e} \right)\qquad \qquad  \qquad E_4 = E_2 + E_D \\
f(E) = \left[1 + \textrm{exp} \left(\frac{E - \mu}{T} \right) \right]^{-1}, \phantom{TEST}
\end{aligned}
\eeq
where at zero temperature, the chemical potential $\mu$ is simply the Fermi energy. In the zero temperature limit, the Fermi-Dirac distribution functions correspond to Heaviside theta functions and so the dynamics structure function becomes
\beq
S(E_D,q) &= \frac{m_e v}{\pi q n_T} \int^\infty_{p_-} dp_2 \ p_2 \,\theta(\mu - E_2) \theta(E_4 - \mu).
\eeq
Writing everything in terms of $p_2$, the two Heaviside theta functions simply become integration limits, so that
\beq
 S(E_D,q) &=& \displaystyle \frac{ m_e v}{\pi q n_T} \displaystyle \int^{\sqrt{2 m_e \mu}}_{\xi} dp_2 \ p_2 \theta(\sqrt{2 m_e \mu} - \xi) \nonumber \\
&=& \displaystyle \frac{ m_e v}{2 \pi q n_T} (2 m_e \mu -\xi^2)  \theta(\sqrt{2 m_e \mu} - \xi),
\eeq
where
\beq
\xi = \textrm{Max}[p_-,\sqrt{2 m_e (\mu-E_D)}].
\eeq

\bibliography{lightdm.bib}
\end{document}